\documentclass[twocolumn,floats,floatfix,prd,superscriptaddress,nofootinbib, amssymb, preprintnumbers]{revtex4-1}

\usepackage{graphicx}
\usepackage{bm}
\usepackage{xcolor}
\usepackage{booktabs}
\PassOptionsToPackage{breaklinks}{hyperref}
\usepackage[linktocpage,colorlinks=true, linkcolor=bluscuro,urlcolor=bluscuro,citecolor=bluscuro]{hyperref}
\definecolor{bluscuro}{rgb}{0.15, 0.2, .85}
\usepackage{amsmath, amssymb}
\usepackage{slashed}
\usepackage{latexsym}
\usepackage{braket}
\usepackage[export]{adjustbox}
\usepackage{color}
\usepackage{multirow}
\usepackage{xspace}
\usepackage{comment}
\usepackage[caption=false]{subfig}
\usepackage{cancel}
\usepackage{url}
\usepackage{nicematrix}

\usepackage[normalem]{ulem} 

\newcommand{\sapienza}{Dipartimento di Fisica, Sapienza Università 
	di Roma, Piazzale Aldo Moro 5, 00185, Roma, Italy}
\newcommand{\infn}{INFN, Sezione di Roma, Piazzale Aldo Moro 2, 00185, Roma, Italy}

\begin{document}

\title{Source multipoles and energy-momentum tensors for spinning black holes\\ and other compact objects in arbitrary dimensions}

\author{Massimo Bianchi}
\email{massimo.bianchi@roma2.infn.it}
\affiliation{Dipartimento di Fisica, Università di Roma ``Tor Vergata'' and Sezione INFN Roma2, Via della Ricerca Scientifica 1, 00133, Roma, Italy}

\author{Claudio Gambino}
\email{claudio.gambino@uniroma1.it}
\affiliation{\sapienza}
\affiliation{\infn}

\author{Paolo Pani}
\email{paolo.pani@uniroma1.it}
\affiliation{\sapienza}
\affiliation{\infn}

\author{Fabio Riccioni}
\email{fabio.riccioni@roma1.infn.it}
\affiliation{\infn}

\begin{abstract}
Working in momentum space and at linear order in the gravitational coupling, we derive the most general class of energy-momentum tensors associated with a given multipolar structure of the spacetime in arbitrary dimensions, and built out of a mass and an angular momentum, at any order in the spin expansion. In this formalism, we are able to derive directly the full multipolar structure of any solution from the multipole expansion of the energy-momentum tensor,  in complete analogy to Newtonian gravity. In particular, we identify the recurrence relations that allow obtaining the multipolar structure of the Kerr and the Myers-Perry black hole solutions, defining source multipoles in a General Relativity context for the first time. For these solutions, we are able to resum the energy-momentum tensor in momentum space at all orders in the angular momentum, and compute its real-space version.  In the Kerr case we exactly obtain the matter source found by Israel, namely an equatorial, pressureless thin disk rotating at superluminal speed.
For Myers-Perry in five dimensions, the matter distribution is a three-ellipsoid in four spatial dimensions with nontrivial stresses.
Remarkably, for any dimensions, the matter configuration is a lower-dimensional distribution  which has the same singularity structure as the fully non-linear black-hole solution.
Our formalism underscores the advantage of working in momentum space to generate nontrivial matter sources for non-linear spacetimes, and could be used to construct regular non-exotic matter configurations that source spinning black hole solutions or horizonless compact objects with the same multipolar structure as black holes. 
\end{abstract}

\maketitle

\tableofcontents

\section{Introduction}\label{sec:Introduction}

The study of black holes~(BHs), particularly rotating ones, has been a cornerstone in understanding the complexities and non-linearities of gravity within the framework of General Relativity. It is hard to overestimate the monumental role that the Kerr solution~\cite{Kerr:1963ud} --~describing the spacetime surrounding a rotating, uncharged BH~-- plays in astrophysics and in high-energy physics.
However, despite its ubiquity across various areas in physics, the precise nature of the matter distribution that generates the Kerr metric remains elusive. This is mainly due to the non-linear nature of the gravitational interaction.

In four spacetime dimensions, this problem has been studied for a long time. 
It is well-known that the matter source of the Schwarzschild metric is a point-like mass located at the origin (which coincides with the singularity of the Schwarzschild spacetime). Furthermore, due to Birkoff's theorem, any spherically symmetric distribution with compact support smaller than the BH horizon would equally source the Schwarzschild solution and might avoid singularities or other pathologies. 
In the spinning case, the situation is more complex.
The Kerr metric can be interpreted as that arising from a material disk that is rotating about its axis of symmetry, rather than from a spinning particle~\cite{Newman:1965tw}. Indeed, in 1970 Israel explicitly computed a source distribution for the Kerr(-Newman~\cite{Newman:1965tw}) metric, in terms of a layer of mass (and charge) distributed over the equatorial disk spanning the ring singularity~\cite{Israel:1970kp}.
Later, making use of the Kerr-Schild gauge, such result has been proved to be mathematically rigorous by using distribution theory to properly deal with the singular source~\cite{Balasin:1993kf}. 
In the uncharged case, this interpretation automatically excises the noncausal parts of the manifold, so that one obtains the source of the causally maximal extension of the vacuum metric.
However, such matter distribution is not physical since the disk rotates at superluminal speed and violates the weak energy condition~\cite{Israel:1970kp}.
There have been various attempts to find non-pathological matter distributions sourcing exactly the Kerr spacetime~\cite{Gurses:1975vu, Lobo:2020ffi,Mazza:2021rgq,Ovalle:2024wtv}. Recently, Ref.~\cite{Maeda:2024tpl} considered a perfect fluid confined inside the horizon that induces a BH geometry which is indistinguishable from the one generated by the Israel matter configuration. However such configurations allow for negative energies violating energy conditions. 
To the best of our knowledge, a matter distribution sourcing the Kerr spacetime and which is free of singularities or pathologies is currently unknown.

In a related context, working in a post-Newtonian approximation, it was recently shown that the infinite tower of  multipole moments of the Kerr spacetime do not uniquely characterize the source: there exist several physically motivated matter configurations with multipole moments identical to those of Kerr~\cite{Bonga:2021ouq}. 
This leaves open the possibility of: i)~finding a regular matter distribution that sources the Kerr spacetime and, ii)~ characterizing BH mimickers~\cite{Cardoso:2019rvt} (\textit{i.e.}, solutions that resemble the BH spacetime at large distance but might be different at the horizon scale) in terms of their multipolar structure. 

Inspired by a quantum field theory approach to General Relativity (GR), in this work we take a new route to the problem of finding the matter source of rotating BHs. We work in momentum space and perturbatively in the gravitational coupling, recasting GR as a quantum field theory for a massless spin-2 field. 
We construct a general class of the energy-momentum tensors (EMTs) describing a spinning point particle in momentum space, to all orders in the particle angular momentum\footnote{Our formalism is general and can be extended to multipole moments that are not induced by angular momentum, possibly breaking equatorial or axial symmetry~\cite{Raposo:2018xkf,Raposo:2020yjy,Bena:2020see,Bianchi:2020bxa,Bena:2020uup,Bianchi:2020miz,Fransen:2022jtw}. For example, it can account for an intrinsic current quadrupole, mass and stress octupoles, or generic moment tensors.}. Since we work at leading order in post-Minkowskian approximation, such construction is fully relativistic and not limited to the description of Newtonian matter configurations. 
After separating the physical terms from the gauge-dependent ones, we find a one-to-one mapping between the angular momentum expansion of the EMT and the (infinite tower of the) spacetime's multipole moments, thus providing a recipe for the class of (momentum-space) EMTs associated with a given multipolar structure. Such direct relation between the source properties and the induced gravitational multipoles is in complete analogy to what one is easily able to do in Newtonian gravity, giving for the first time a simple recipe to read the multipole moments without computing the metric as an intermediate step. Such approach leads to an unambiguous definition of source multipole moments in a relativistic context, generalizing the work in~\cite{Bonga:2021ouq}.

As we will discuss, this approach has several advantages. First, the framework can be naturally developed in arbitrary dimensions, which allows us to discuss both the Kerr case and its higher-dimensional generalization, the Myers-Perry solution~\cite{Myers:1986un}, and to connect with the new stress multipole moments that exist in spacetime dimensions higher than four~\cite{Gambino:2024uge, Heynen:2023sin}.
Second, working in momentum space allows us to clearly distinguish between local and non-local contributions (to be rigorously defined later), with only the latter affecting the multipole moments. 
This identification is crucial to obtain a tractable angular-momentum expansion. 
Remarkably, for Myers-Perry BHs in generic spacetime dimensions (including Kerr in four dimensions), such an expansion of the EMT can be resummed in momentum space and, perhaps even more surprisingly, the resummed stress-energy tensor can be analytically transformed to coordinate space.
We therefore obtain a general expression for the EMT sourcing a spinning BH (with spherical topology) in arbitrary dimensions.

In the Kerr case this EMT exactly corresponds to the matter source found by Israel, namely a thin disk rotating at superluminal speed, whereas for Myers-Perry in five spacetime dimensions the matter distribution
is a three-dimensional ellipsoid with nontrivial stresses in four spacial dimensions.
In both cases the matter distribution lives in lower dimensions than the fully non-linear BH solution, but it shares the same singularity structure. 

This is a remarkable property emerging from our approach: by developing the most general field theory for a monopole and a dipole \emph{in momentum space} in arbitrary dimensions and resumming the result to all orders in angular momentum, we obtain a nontrivial extended (but localized) matter distribution in coordinate space, which is exactly the one sourcing the same multipolar structure as a (non-linear) spinning BH spacetime. Interestingly, this already happens at first order in the coupling constant $G$. 
This is a rather surprising result, as it suggests that the characteristic curvature singularity of BHs may not be non-linear in nature, since it appears already at the leading order in $G$. Indeed, since the result is derived simply by imposing the BH multipole structure, it indicates a strong connection between infrared and ultraviolet BH physics.

The paper is organized as follows. In Sec.~\ref{sec:EMTcalc} we define the most generic spin-induced energy-momentum tensor up to local contributions, identifying physical and gauge degrees of freedom. In Sec.~\ref{Sec:GravMultFFO} we relate the physical terms to the multipolar structure of the corresponding spacetime and in Sec.~\ref{sec:ATSBHs} we apply such formalism to the case of black holes in arbitrary dimensions. Then in Sec.~\ref{sec:MSKMP} we determine the source of such black holes by resumming over the spin the energy-momentum tensor in momentum space. Finally Sec.~\ref{sec:Conclusions} contains our conclusions. 

{\bf Conventions.}
We work in the mostly positive signature with $\eta_{00}=-1$ and in natural units, $\hbar=c=1$, whereas we keep the gravitational coupling constant $G$ explicit. Greek indices are for spacetime components $\mu,\nu=0, 1, ..., d$ and Latin indices are for space components only $i, j=1, ..., d$, where $D=d+1$ is the number of spacetime dimensions.

\section{Spin-induced energy-momentum tensor in momentum space}\label{sec:EMTcalc}
We consider GR in arbitrary dimensions,
\begin{equation}
    G_{\mu\nu}=8\pi G T_{\mu\nu}\,,
\end{equation}
and a linearized gravity framework, 
\begin{equation} \label{eq:expansioninh}
    g_{\mu\nu}=\eta_{\mu\nu}+\kappa \, h_{\mu\nu} + O(G^2)\ ,
\end{equation}
where $\kappa^2=32\pi G$ and we adopt the convention that $h_{\mu\nu}$ itself is linear in $\kappa$, which is commonly used in scattering amplitude calculations involving gravitons. After fixing the gauge and solving the Einstein equations at ${\cal O}(G)$, one gets 
\begin{equation}
    h_{\mu\nu}(t, \vec{x})=\frac{\kappa}{2}\int \frac{d^{d+1} q}{(2\pi)^{d+1}}\frac{e^{+i tq^0-i \vec q\cdot \vec x}}{q^2}P_{\mu\nu, \rho \sigma}T^{\rho\sigma}(q)\ ,
\end{equation}
where $P_{\mu\nu, \rho\sigma}$ is the transverse projector in the propagator in some gauge. Restricting our attention to stationary solutions, from here on we will always consider
\begin{equation}
    T^{\rho\sigma}(q)\rightarrow 2\pi\delta(q^0)T^{\rho\sigma}(\vec q\, )\ ,
\end{equation}
such that $q^2=\vec q\, ^2$ inside the integral and the metric simplifies to the form
\begin{equation}\label{eq:MetricFromEMT}
    h_{\mu\nu}(\vec{x})=\frac{\kappa}{2}\int \frac{d^d q}{(2\pi)^d}\frac{e^{-i \vec q\cdot \vec x}}{q^2}P_{\mu\nu, \rho \sigma}T^{\rho\sigma}(\vec q\, )\ .
\end{equation}
The metric perturbation then is written in terms of the Fourier transform of the EMT, which is defined by
\begin{equation}\label{eq:MomentumEMT}
    T_{\mu\nu}(\vec{x})=\int \frac{d^d q}{(2\pi)^d}e^{-i \vec q\cdot \vec x}\ T_{\mu\nu}(\vec q\, )\ .
\end{equation}
Within linearized gravity, both the EMT and the metric perturbation $h_{\mu\nu}$ are tensors living in a flat background, and so the indices are raised and lowered by the Minkowski metric. 
From now on, for ease of notation, we will omit the vector symbol in the argument of the tensor fields, \textit{i.e.} we write
$h_{\mu\nu}({x})\equiv h_{\mu\nu}(\vec{x})$, $T_{\mu\nu}({q})\equiv T_{\mu\nu}(\vec{q}\, )$, and so on.
\vspace{0.5cm}

Our goal here is to derive a general description for a rotating spin-induced source, namely the EMT in momentum space produced when the only scales (and physical objects) involved are the ADM mass $m$ and angular momentum tensor, whose magnitude is $J$. To do so, we get inspiration from the EMT in momentum space at quadrupole order associated to a spin-1 massive particle recovered in~\cite{Gambino:2024uge} from a scattering amplitude approach. Such result, can be generalized to every order in the spin expansion just by taking into account every possible tensorial structure, while satisfying some requirements that we are going to discuss in the following.
\vspace{0.5cm}

Defining $u^\mu$ and $J^{\mu\nu}$ as the velocity and the anti-symmetric spin tensor of the EMT, in the stationary case it follows that $J^{\mu\nu}u_\nu=0$ and $q^\mu u_{\mu}=0$. In the simple case in which the source is sitting at the origin one has $u^0=-1$ and $u^i=0$. Then, considering as usual the angular momentum per unit mass $J/m$, the only combination of variables we have at our disposal to build the generic expression of the EMT in momentum space is $q J/m$, since we require a smooth limit in which $J/m\rightarrow 0$ and $m\rightarrow 0$. The empiric rule then, is that for every spin tensor $J$ there must be a transferred momentum $q$.
\vspace{0.5cm}

Finally, we restrict to the study of the long-range regime of the metric in Eq.~\eqref{eq:MetricFromEMT}, and so we neglect terms in which $T_{\mu\nu}(q)\propto q^2$, that otherwise will give rise to local contributions (discussed in detail in Sec.~\ref{sec:LocalContributions})\footnote{Henceforth, short- or long-range contributions will always refer to the metric behavior.}. Indeed, such local contributions would cancel the graviton propagator in Eq.~\eqref{eq:MetricFromEMT}, leading to a term in the metric perturbation proportional to a delta function (or derivatives of delta functions). To be more precise, using the short-hand notation
\begin{equation}
    q\cdot J\cdot J\cdot q \equiv q^\mu J_{\mu}{}^{\nu}J_{\nu}{}^{\sigma}q_{\sigma}\ ,\quad J\cdot J=J^{\mu\nu}J_{\nu\mu}\ ,
\end{equation}
terms like
\begin{equation}
    T^{\mu\nu}(q)\propto u^{\mu}u^{\nu} \frac{q^2}{m^2} J \cdot J
\end{equation}
lead to local contributions in the metric, as in 
\begin{equation}\label{eq:DeltaMetric}
    h_{\mu\nu}(x)\propto \delta(x)\ .
\end{equation}
On the other hand, terms like 
\begin{equation}
    T^{\mu\nu}(q)\propto u^{\mu}u^{\nu} \frac{q\cdot J \cdot J \cdot q}{m^2}\ ,
\end{equation}
are associated with non-local contributions in the metric, as in 
\begin{equation}
    h_{\mu\nu}(x)\propto \frac{1}{r}\ .
\end{equation}

Within the aforementioned assumptions, the most generic expression of the EMT reads
\begin{widetext}
\begin{equation}\label{eq:GenericEMT}
    \begin{aligned}
    &T^{\mu\nu}(q)=m\ u^{\mu}u^{\nu}\Bigg(1+\sum_{n=1}^{+\infty}{F}_{2n, 1}\left(-\frac{q\cdot J\cdot J\cdot q}{m^2}\right)^n\Bigg)+m\sum_{n=0}^{+\infty}{F}_{2n+2, 2}\frac{(J\cdot q)^\mu (J\cdot q)^\nu}{m^2}\left(-\frac{q\cdot J\cdot J\cdot q}{m^2}\right)^n\\
    &-\frac{i}{2}m\left(u^\mu\frac{(J\cdot q)^\nu}{m} +u^\nu\frac{(J\cdot q)^\mu}{m} \right)\Bigg(1+\sum_{n=1}^{+\infty}{F}_{2n+1, 3}\left(-\frac{q\cdot J\cdot J\cdot q}{m^2}\right)^n\Bigg)\\
    &-m\sum_{n=0}^{+\infty}{G}_{2n+2, 1}\Bigg(\eta^{\mu\nu}\frac{q\cdot J\cdot J\cdot q}{m^2}-\frac{(J\cdot J\cdot q)^\mu q^\nu+(J\cdot J\cdot q)^\nu q^\mu}{m^2}\Bigg)\left(-\frac{q\cdot J\cdot J\cdot q}{m^2}\right)^n\\
    &-m\sum_{n=0}^{+\infty}{G}_{2n+2, 2}\frac{q^\mu q^\nu }{m^2}J\cdot J\left(-\frac{q\cdot J\cdot J\cdot q}{m^2}\right)^n+m\sum_{n=0}^{+\infty}{G}_{2n+4, 3}\frac{q^\mu q^\nu}{m^2}\frac{q\cdot J\cdot J\cdot J\cdot J\cdot q}{m^2}\left(-\frac{q\cdot J\cdot J\cdot q}{m^2}\right)^n\ .
    \end{aligned}
\end{equation}
\end{widetext}
We stress that every other terms other than the ones in Eq.~\eqref{eq:GenericEMT} are local terms.

Eq.~\eqref{eq:GenericEMT} naturally organizes in a spin expansion of the source, where the $F_{n, m}$'s and $G_{n, m}$'s are constant terms, with the first index labelling the order of the spin to which each coefficient is referring to, and the second one labelling  different coefficients. Their distinct nature will be discussed in detail later. Moreover, the mass and the angular momenta are normalized to the ADM values of the induced spacetime, meaning that, computing the linearized metric from Eq.~\eqref{eq:GenericEMT}, the monopole and the dipole terms are fixed in terms of $m$ and $J$ by setting $F_{0, 1}=F_{1, 3}=1$.

It is possible to check that the EMT in Eq.~\eqref{eq:GenericEMT} is conserved up to local terms, namely $q_\mu T^{\mu\nu}(q)\propto  q^2$, and it reduces to a point-like mass $m$ sitting at the origin in the spinless limit. Notice that neglecting local terms is a crucial ingredient to obtain a finite number of terms that describe the EMT at every spin-order, as well as working in momentum space. Indeed, the same argument could not be repeated in position space, in which even restricting to the long-range regime an infinite number of tensorial structures is needed in order to capture the full spin-expansion in arbitrary dimension.   

\subsection{Gauge redundant parameters}\label{sec:GaugeTransformation}

Considering Eq.~\eqref{eq:GenericEMT}, we can ask ourselves if every term in the EMT is physical. Indeed, we can fix a gauge in which we evaluate the metric induced by the source and perform a coordinate transformation eliminating any potentially redundant gauge parameters.  
As a standard choice we fix the harmonic gauge
\begin{equation}
g^{\mu\nu}\Gamma_{\mu\nu}^{\alpha}=\kappa\  \partial_{\mu}\left(h^{\mu \alpha}-\frac{1}{2}\eta^{\mu \alpha}h\right)+O(G^2)=0\ ,
\end{equation}
in which the relevant projector reads
\begin{equation}
    P_{\mu\nu, \rho \sigma}=\frac{1}{2}\left(\eta_{\mu\rho}\eta_{\nu\sigma}+\eta_{\mu\sigma}\eta_{\nu\rho}-\frac{2}{d-1}\eta_{\mu\nu}\eta_{\rho\sigma}\right)\ .
\end{equation}
This gauge choice allows us to specialize Eq.~\eqref{eq:MetricFromEMT} to the form
\begin{equation}\label{eq:MetricFT}
    h_{\mu\nu}(x)=\frac{\kappa}{2}\int \frac{d^d q}{(2\pi)^d}\frac{e^{-i \vec q\cdot \vec x}}{q^2}\left(T_{\mu\nu}(q)-\frac{1}{d-1}\eta_{\mu\nu}T(q)\right)\ ,
\end{equation}
where $T(q)=\eta_{\mu\nu}T^{\mu\nu}(q)$. 

Consider now an infinitesimal coordinate transformation $x'_\mu=x_\mu+\xi_\mu(x)$ such that
\begin{equation}\label{eq:ShiftedMetric}
    \kappa\, h'_{\mu\nu}=\kappa\, h_{\mu\nu}-(\partial_\mu\xi_\nu+\partial_\nu\xi_\mu)\ .
\end{equation}
At infinity we can reduce to vacuum Einstein equations, such that $\Box h_{\mu\nu}(x)=0$.
Since in the harmonic gauge ${\Box x_\mu=\Box x'_\mu=0}$, every coordinate transformation satisfying $\Box \xi_\mu=0$ would preserve this gauge. We can then express the shift in terms of the generalized harmonic function
\begin{equation}\label{Eq:RhoDefinition}
    \rho=\frac{\Gamma(d/2-1)\pi^{1-d/2}}{r^{d-2}}\ ,
\end{equation}
such that $\Box \rho=0$. Then, fixing $\xi^0=0$, and defining the angular momentum density tensor
\begin{equation}
    S^{\mu\nu}=J^{\mu\nu}/m\ ,
\end{equation}
the most general gauge shift can be expressed as
\begin{equation}\label{eq:CoordTransf}
\begin{aligned}
    &\xi^i(x)=\frac{\kappa^2m}{8\pi}\sum_{n=0}^{+\infty}\Bigg(K_{2n+2, 1}(S\cdot S)^{iA_{2n+1}}\\
    &+K_{2n+2, 2}(S\cdot S) \eta^{ia_1}(S\cdot S)^{A_{2n}}\Bigg)\partial_{A_{2n+1}}\rho\\
    &+\frac{\kappa^2m}{8\pi}\sum_{n=0}^{+\infty}\Bigg(K_{2n+4, 3}(S\cdot S\cdot S\cdot S)^{a_1a_2}\eta^{ia_3}(S\cdot S)^{A_{2n}}\Bigg)\partial_{A_{2n+3}}\rho\ ,
\end{aligned}
\end{equation}
where we used the short-hand notation 
\begin{equation}
    (S\cdot S)^{A_{2n}}\partial_{A_{2n}}\rho\equiv (S\cdot S)^{a_1a_2}\cdots(S\cdot S)^{a_{n-1}a_n}\partial_{a_1}\cdots\partial_{a_n}\rho\ .
\end{equation}

Finally, computing the metric in harmonic gauge and considering the shift in Eq.~\eqref{eq:ShiftedMetric} as shown in Appendix \ref{App:MetricComp}, fixing 
\begin{equation}\label{eq:GaugeFixingFFF}
\begin{gathered}
    K_{2n+2, 1}=-G_{2n+2,1}\ , \qquad K_{2n+2, 2}=\frac{1}{2}G_{2n+2, 2}\ ,\\ K_{2n+4, 3}=\frac{1}{2}G_{2n+4, 3}\ ,
\end{gathered}
\end{equation}
we can completely eliminate the dependence on the $G_{n, m}$'s from the metric, leaving it only dependent on the $F_{n, m}$'s. Since the latter are the only terms that cannot be canceled by a coordinate transformation, they have a physical meaning. For this reason, we dub the $G_{n, m}$ and the $F_{n, m}$ coefficients as ``residual factors'' and ``form factors'', respectively.

As we shall discuss in the next section, form factors are tightly related to gravitational multipoles, and they can be mapped into each other by using the higher dimensional generalization of Thorne formalism~\cite{Thorne:1980ru, Heynen:2023sin,Gambino:2024uge}. Indeed, the number of independent towers of form factors is related to the fact that in higher dimensions there exist three different towers of multipoles, namely mass, current, and stress multipole moments~\cite{Gambino:2024uge}, 
and the latter can be gauged away only in $d=3$.
Thus, for spin-induced multipoles there are two physical degrees of freedom (the mass and stress) for each even-power spin term, and one degree of freedom (the current) for each odd-power contribution.

\section{Gravitational multipoles from form factors}\label{Sec:GravMultFFO}

Let us consider the multipole expansion of the linearized metric in ACMC coordinates within the generalized Thorne formalism~\cite{Gambino:2024uge} with the following normalization
\begin{equation}\label{eq:MultipoleExpandedMetric}
\begin{aligned}
    g_{00}&=-1+4\frac{d-2}{d-1}\sum_{\ell=0}^{+\infty}\frac{G m \rho(r)}{r^\ell}\mathbb{M}^{(\ell)}_{A_\ell}N_{A_\ell}+\cdots\ ,\\
    g_{0i}&=2(d-2)\sum_{\ell=1}^{+\infty}\frac{Gm\rho(r)}{r^\ell}\mathbb{J}^{(\ell)}_{i, A_\ell}N_{A_\ell}+\cdots\ ,\\ 
    g_{ij}&=\delta_{ij}+4\frac{d-2}{d-1}\sum_{\ell=2}^{+\infty}\frac{G m \rho(r)}{r^\ell}\tilde{\mathbb{G}}^{(\ell)}_{ij,A_\ell}N_{A_\ell}+\cdots\ ,
\end{aligned}
\end{equation}
where $A_\ell=a_1\cdots a_\ell$,  $N_{A_\ell}=\frac{x_{a_1}\cdots \,x_{a_\ell}}{r^\ell}$,  $\mathbb{M}^{(\ell)}_{A_\ell}$ and $\mathbb{J}^{(\ell)}_{i, A_\ell}$ are the mass and current multipole moments, respectively, and $\tilde{\mathbb{G}}^{(\ell)}_{ij,A_\ell}$ are related to the stress multipoles through the following relation 
\begin{equation}
\mathbb{G}^{(\ell)}_{ij,A_\ell}=\tilde{\mathbb{G}}^{(\ell)}_{ij,A_\ell}+\frac{1}{2}\delta_{ij}\Big(\mathbb{M}^{(\ell)}_{A_\ell}-\tilde{\mathbb{G}}^{(\ell)}_{kk,A_\ell}\Big)\ ,
\end{equation}
with $\mathbb{G}^{(\ell)}_{ij,A_\ell}$ being the stress multipoles defined in~\cite{Gambino:2024uge} (see also~\cite{Amalberti:2023ohj}). 
Having normalized the mass $m$ and the spin density $S$ to their ADM value in Eq.~\eqref{eq:GenericEMT}, results in having fixed the mass monopole and the spin dipole to the specific value of 
\begin{equation}
    \mathbb{M}^{(0)}=1\ , \qquad \mathbb{J}^{(1)}_{i a_1}=S_{ia_1}\ ,
\end{equation}
while one can impose without loss of generality that
 \begin{equation}
    \mathbb{M}^{(1)}=0 \ , \qquad \mathbb{G}_{ij}^{(0)}=0\ , \qquad \mathbb{G}_{ij}^{(1)}=0\ .
\end{equation}

We can now compare the generalized multipole expansion of the metric in \eqref{eq:MultipoleExpandedMetric} with the generic metric sourced by the EMT in Eq.~\eqref{eq:GenericEMT} and shown in Appendix \ref{App:MetricComp},
thus establishing a relation between gravitational multipoles and the source form factors. Indeed, such metric is expressed in harmonic coordinates which, for vacuum spacetimes, are a particular choice of the ACMC gauge~\cite{Mayerson:2022ekj}.
The result then reads
\begin{widetext}
\begin{equation}\label{eq:GravitationalMultipoles}
    \begin{aligned}
\mathbb{M}^{(2\ell)}_{A_{2\ell}}&=\frac{(d+4\ell-4)!!}{(d-2)!!}(-1)^\ell\Big(F_{2\ell, 2}+(d-2)F_{2\ell, 1}\Big)(-S\cdot S)_{A_{2\ell}}\Big|_{\rm STF}\ , \\
\mathbb{J}^{(2\ell+1)}_{i,A_{2\ell+1}}&=\frac{(d+4\ell-2)!!}{(d-2)!!}(-1)^\ell F_{2\ell+1, 3} \ S_{ia_1}(-S\cdot S)_{A_{2\ell}}|_{\rm ASTF}\ , \\
\mathbb{G}^{(2\ell)}_{ij,A_{2\ell}}&=(d-1)\frac{(d+4\ell-4)!!}{(d-2)!!}(-1)^\ell F_{2\ell, 2} \ S_{ia_1}S_{ja_2}(-S\cdot S)_{A_{2\ell-2}}|_{\rm RSTF}\ ,
    \end{aligned}
\end{equation}
\end{widetext}
with 
\begin{equation}
      \mathbb{M}^{(2\ell+1)}_{A_{2\ell+1}}=0\ ,\qquad \mathbb{J}^{(2\ell)}_{i, A_{2\ell}}=0\ ,\qquad\mathbb{G}^{(2\ell+1)}_{ij, A_{2\ell+1}}=0\ ,
\end{equation}
for every $\ell=0, 1, 2, \dots$. As expected in analogy with the $d=3$ case, mass and stress multipoles are non-vanishing only for even powers of the spin, while current multipoles are non-vanishing for odd powers. Indeed, every multipole tensor in Eq.~\eqref{eq:GravitationalMultipoles} is meant to be symmetrized in the correct way, even though once contracted with $N_{A_{\ell}}$ the symmetrization is not relevant. In particular, the mass multipole must be STF (symmetric and trace-free)~\cite{Thorne:1980ru}, the current multipole ASTF~\cite{Heynen:2023sin}, meaning that is STF with respect $A_{2\ell}$, anti-symmetric in $(ia_1)$, traceless with respect all indices and with any antisymmetrization over three indices vanishing, and the stress multipole must be RSTF (Riemann-symmetric and trace-free)~\cite{Gambino:2024uge}, namely STF in the $A_{2\ell-2}$, respecting the symmetries of the Riemann tensor for every $(ia_1,ja_2)$ (including the Bianchi identity), traceless with respect to all indices and with any antisymmetrization of three indices vanishing. Moreover, since the multipole moments defined in Eq.~\eqref{eq:GravitationalMultipoles} only depend on the form factors, we notice that the argument in Sec.~\ref{sec:GaugeTransformation} shows that they are not gauge dependent, since a coordinate transformation only affects residual factors. 

Since we are considering spin-induced multipole moments, the tensors in Eq.~\eqref{eq:GravitationalMultipoles} are axially-symmetric around each rotational plane, and they are also symmetric with respect to the exchange of every pair of angular momenta\footnote{In $d=4, 5$ with two distinguished angular momenta, this corresponds to a bi-axial symmetry, in $d=6,7$ to a tri-axial symmetry, and so on.}. These symmetries drastically reduce the number of independent components of each multipole moment, such that for instance in $d=3, 4$ the mass moments have only a single component. However, in higher dimensions this number grows, as well as the number of independent components of the current and stress moments. Hence, using Cartesian coordinates and expressing the multipole towers in terms of tensors like in \eqref{eq:GravitationalMultipoles}, is the simplest way to deal with multipoles in higher dimensions. It is important to stress that the gauge invariance of Eq.~\eqref{eq:GravitationalMultipoles} under ACMC preserving coordinate transformations was proved only in $d=3$~\cite{Geroch:1970cc, Geroch:1970cd, Hansen:1974zz, Gursel1983}, and for the mass and current multipoles in $d=4$~\cite{Heynen:2023sin}. Based on this, we conjecture that Eq.~\eqref{eq:GravitationalMultipoles} be gauge invariant in any dimensions.

One of the main results that can be read off from Eq.~\eqref{eq:GravitationalMultipoles} is that, in the spin induced case, we are able to connect directly the source form factors to the gravitational multipoles. Using this recipe, one does not need to compute the metric induced by some EMT in order to find the multipole moments, but one just needs to know the form factors of the source, precisely as one would do in Newtonian gravity~\cite{Bonga:2021ouq}. This relation is the relativistic analog of the correspondence between source multipoles and Newtonian gravitational multipoles for non-relativistic systems, and in this sense form factors can be seen as an unambiguous definition of source multipoles in GR, generalizing such source-gravity multipole relation for relativistic theories for the first time. Even if Eq.~\eqref{eq:GravitationalMultipoles} is specialized for spin-induced moments, we strongly believe that such duality can be established also for generic multipole moments by extending our formalism to other fundamental multipole moments \cite{Raposo:2018xkf}. 

Later on we will match the metric in \eqref{eq:MetricFromEMT} with known GR solutions, in particular we will consider the Kerr metric and the Myers-Perry BH in $d=4$, in order to obtain the relative form factors, and hence the gravitational multipoles of such solutions, eventually generalizing the results to arbitrary dimensions. To this end, it is worth seeing in more details the main differences in Eq.~\eqref{eq:GravitationalMultipoles} between the $d=3$ and its higher dimensional version. 

As already discussed in~\cite{Gambino:2024uge}, $d=3$ is a particularly special case, since in $SO(3)$ the dimension of the fundamental representation coincides with the adjoint one, and so an anti-symmetric rank-2 tensor is dual to a vector. This results in the possibility to define a spin vector as 
\begin{equation}\label{eq:SpinVector}
    S^{ij}=\varepsilon^{ijk}s_k\ ,
\end{equation}
where $\varepsilon^{ijk}$ is the Levi-Civita symbol and $s_k$ the spin vector. Replacing Eq.~\eqref{eq:SpinVector} into \eqref{eq:GravitationalMultipoles} one gets
\begin{equation}\label{eq:d3Multipoles}
    \begin{aligned}
        \mathbb{M}^{(2\ell)}_{A_{2\ell}}\Big|_{d=3}&=(4\ell-1)!!\Big(F_{2\ell, 1}+F_{2\ell, 2}\Big)s_{a_1}\cdots s_{a_{2\ell}}\Big|_{TF}\ , \\         \mathbb{J}^{(2\ell+1)}_{i,A_{2\ell+1}}\Big|_{d=3}&=(4\ell)!!F_{2\ell+1, 3} \ \epsilon_{ia_1k}s_{k}s_{a_2}\cdots s_{a_{2\ell}}\Big|_{TF}\ , \\
         \mathbb{G}^{(2\ell)}_{ij,A_{2\ell}}\Big|_{d=3}&=0\ ,
    \end{aligned}
\end{equation}
where the tensors are meant to be made trace-free (TF). The first thing to be noticed is that, differently from the higher dimensional case, in $d=3$ the multipoles can be expressed in terms of just STF tensors, namely $s_{A_{\ell}}|_{\rm STF}=s_{a_1}\cdots s_{a_{\ell}}|_{TF}$. Furthermore, as noticed in~\cite{Gambino:2024uge}, the stress multipoles vanish and the form factor $F_{2\ell, 2}$ is redundant. This means that only the combination $F_{2\ell, 1}+F_{2\ell, 2}$ is physical and we have more gauge freedom to compute the EMT in \eqref{eq:GenericEMT}. In order to explicitly see how the stress multipole tensor vanishes in $d=3$ let us discuss the quadrupole case. In arbitrary dimension, the explicit RSTF projection of the quadrupole tensor structure reads
\begin{equation}
\begin{aligned}
   &S_{ia_1}S_{ja_2}\Big|_{\rm RSTF}=S_{ia_1}S_{ja_2}-\frac{1}{3}\Big(S_{ia_1}S_{ja_2}+S_{a_1j}S_{ia_2}+S_{ji}S_{a_1a_2}\Big)\\
   &+\frac{1}{d-2}\Big(S_{a_1k}S^{k}{}_{a_2}\delta_{ij}-S_{a_1k}S^{k}{}_{j}\delta_{ia_2}-S_{ik}S^{k}{}_{a_2}\delta_{a_1j}+S_{ik}S^{k}{}_{j}\delta_{a_1a_2}\Big)\\
   &+\frac{1}{(d-2)(d-1)}\Big(S_{k_1k_2}S^{k_2k_1}\delta_{ia_2}\delta_{ja_1}-S_{k_1k_2}S^{k_2k_1}\delta_{ij}\delta_{a_1a_2}\Big)\ ,
\end{aligned}
\end{equation}
respecting manifestly all the RSTF symmetries. It is then easy  to see that for $d=3$, replacing Eq.~\eqref{eq:SpinVector} and expressing the product of the Levi-Civita symbols in terms of delta's, one gets exactly
\begin{equation}
    S_{ia_1}S_{ja_2}\Big|^{d=3}_{\rm RSTF}=0\ .
\end{equation}

\subsection{Conserved EMT in momentum space} \label{sec:ConservedEMT}
Even though Eq.~\eqref{eq:GenericEMT} is enough to reconstruct the long-range behavior of the metric induced by the source, \textit{i.e.} the multipole moments, if we are interested in computing the EMT in position space we have to make sure that it is properly conserved, including also possible local-term contributions. We can indeed consider the properly conserved EMT as
\begin{widetext}
\begin{equation}\label{eq:ConservedEMT}
    \begin{aligned}
    &T^{\mu\nu}(q)=m\ u^{\mu}u^{\nu}\Bigg(1+\sum_{n=1}^{+\infty}{F}_{2n, 1}\Big(-q\cdot S\cdot S\cdot q\Big)^n\Bigg)+m\sum_{n=0}^{+\infty}{F}_{2n+2, 2}(S\cdot q)^\mu (S\cdot q)^\nu\Big(-q\cdot S\cdot S\cdot q\Big)^n\\
    &+m\sum_{n=0}^{+\infty}{G}_{2n+2, 1}\Bigg(\eta^{\mu\nu}\Big(-q\cdot S\cdot S\cdot q\Big)-(S\cdot S)^{\mu\nu}q^2+(S\cdot S\cdot q)^\mu q^\nu+(S\cdot S\cdot q)^\nu q^\mu\Bigg)\Big(-q\cdot S\cdot S\cdot q\Big)^n\\
    &+m\sum_{n=0}^{+\infty}{G}_{2n+2, 2}\Big(q^\mu q^\nu-\eta^{\mu\nu}q^2\Big)(-S\cdot S)\Big(-q\cdot S\cdot S\cdot q\Big)^n+m\sum_{n=0}^{+\infty}{G}_{2n+4, 3}\Big(q^\mu q^\nu-\eta^{\mu\nu}q^2\Big)\Big(q\cdot S\cdot S\cdot S\cdot S\cdot q\Big)\Big(-q\cdot S\cdot S\cdot q\Big)^n\\
    &-\frac{i}{2}m\Big(u^\mu(S\cdot q)^\nu +u^\nu(S\cdot q)^\mu \Big)\Bigg(1+\sum_{n=1}^{+\infty}{F}_{2n+1, 3}\Big(-q\cdot S\cdot S\cdot q\Big)^n\Bigg)\ ,
    \end{aligned}
\end{equation}
\end{widetext}
where now $q_\mu T^{\mu\nu}(q)=0$, as requested by the fact that the EMT in coordinate space is divergence-free. 
Since the only difference with 
respect to \eqref{eq:GenericEMT} is due to terms depending on $q^2$, Eq.~\eqref{eq:ConservedEMT} gives rise to the same long-range behavior of the metric.  

When including local terms in Eq.~\eqref{eq:ConservedEMT} we have to be careful with the residual factors $G_{n, m}$, and in particular check whether or not they are still redundant parameters. In Sec.~\ref{sec:GaugeTransformation} we have proved that they do not physically contribute to the asymptotic behavior of the metric; however, they do give rise to local contributions in the gravitational field. This means that if one is interested in computing the coordinate-space version of the EMT, or the short-range behavior of the metric, these terms are not negligible and have to be taken into account. 

Indeed, once the form factors are fixed, there is an infinite number of different sources that reproduce the exact same multipolar structure and differ among each other only by local contributions, both in the EMT and in the metric. Moreover, allowing for local contributions, Eq.~\eqref{eq:ConservedEMT} is no more the most generic EMT in momentum space, since there are now infinite terms (tensorial structures) contributing in such regime. Still, we can think of Eq.~\eqref{eq:ConservedEMT} as the most generic EMT modulo local terms that do not affect the multipolar structure.

\subsection{Local contributions}\label{sec:LocalContributions}

As we already pointed out, in our construction of Eq.~\eqref{eq:GenericEMT} local terms are neglected, while in Eq.~\eqref{eq:ConservedEMT} we have included them to enforce the conservation of the EMT. However, even though they do not contribute to gravitational multipoles, they are important to characterize an equivalence class of sources that share the same multipolar structure. 
We have two ways to account for local terms in the EMT: i)~by considering more tensorial structures in Eq.~\eqref{eq:ConservedEMT} that are purely local; ii)~by promoting the form factors or the residual factors (or also any extra purely local term coefficients) to analytic functions of $q^2$ by adding one or more new scales to the source. 

To clarify how such local terms modify the gravitational field sourced by the EMT let us study a simple example. Consider a matter-energy distribution with only a non-vanishing monopole,
\begin{equation}
    T_{\mu\nu}(q)=m \, u_\mu u_\nu F_{0, 1}\ ,
\end{equation}
and promote the form factor to an analytic function of $q^2$ such that it maintains the ADM normalization of the mass, for instance
\begin{equation}\label{eq:PromotionFF}
    F_{0, 1}\rightarrow F_{0, 1}(q^2)=\frac{3}{q^3\mathcal{R}^3}\Big(\sin(q\mathcal{R})-q\mathcal{R}\cos(q\mathcal{R})\Big)\ ,
\end{equation}
where $\mathcal{R}$ is a length scale characterizing the size of the source. Expanding Eq.~\eqref{eq:PromotionFF} we can see that it behaves exactly as we requested
\begin{equation}\label{eq:EMTdustExpanded}
    F_{0, 1}(q^2)=1-\frac{1}{10}\mathcal{R}^2 q^2+\frac{1}{280}\mathcal{R}^4 q^4+O(\mathcal{R}^6q^6)\ ,
\end{equation}
meaning that $F_{0, 1}(0)=1$ (leaving the mass normalized at its ADM value) and it is an analytic function of $q^2$. 

Computing the Fourier transform in $d=3$ of such EMT one gets
\begin{equation}\label{eq:dustEMT}
    T_{\mu\nu}(x)= \epsilon \ u_\mu u_\nu\   \Theta (\mathcal{R}-r)\ ,
\end{equation}
where $r$ is the radial coordinate, $\Theta(x)$ is the Heaviside step function, and 
\begin{equation}
    \epsilon=\frac{m}{\frac{4}{3}\pi \mathcal{R}^3}\ .
\end{equation}
Hence, Eq.~\eqref{eq:dustEMT} just describes a spherical pressureless dust of radius $\mathcal{R}$ and constant density $\epsilon$. Finally, we can compute the linearized metric sourced by such EMT through Eq.~\eqref{eq:MetricFromEMT} as
\begin{equation}\label{eq:dustMetric}
    \kappa\, h_{00}=\begin{cases} \frac{G m(r)}{r}\left(3\frac{\mathcal{R}^2}{r^2}-1\right) \quad &r<\mathcal{R}\\
    \frac{2 Gm}{r}\quad &r>\mathcal{R}
    \end{cases}\ ,
\end{equation}
with $m(r)$ the radial mass function defined by 
\begin{equation}
    m(r)=\frac{4}{3}\pi r^3\epsilon \ \Theta(\mathcal{R}-r)+m\ \Theta(r-\mathcal{R})\ .
\end{equation}

From Eqs.~\eqref{eq:EMTdustExpanded} and \eqref{eq:dustMetric}, it is evident that the leading-order term of the promoted monopole form factor fully reconstructs the long-distance behavior of the metric for $r > \mathcal{R}$. In contrast, the series of local terms—when expanded order by order in $\mathcal{R}$—would correspond to delta function contributions to the metric, as shown in Eq.~\eqref{eq:DeltaMetric}. However, after resummation, these terms collectively describe the gravitational field for $r < \mathcal{R}$, effectively smoothing out the singularity at $r = 0$.

We can further complicate the previous example to have a richer phenomenology even in the spherically-symmetric case. Since we introduced a new length scale in the system, we can write down a residual factor contributing at zero-order in the spin, and consider for example the most generic analytic spherically-symmetric EMT in momentum space described only by $m$ and $\mathcal{R}$ as
\begin{equation}\label{eq:EMTsfComplicated}
    T_{\mu\nu}(q)=mu_\mu u_\nu F_{0, 1}(q^2)+m \mathcal{R}^2(q_\mu q_\nu-\eta_{\mu\nu}q^2) G_0(q^2)\ ,
\end{equation}
where $G_0(q^2)$ is the residual factor at zero-order in the spin, already promoted to an analytic function of $q^2$.
In Eq.~\eqref{eq:EMTsfComplicated} the term proportional to $q_\mu q_\nu$ can be gauged away from the metric with a coordinate transformation similar to Eq.~\eqref{eq:CoordTransf}, while the local term will modify both the EMT and the metric when $r\sim \mathcal{R}$. 

Generalizing the above argument, considering $\mathcal{R}_n$ the set of all the scales that characterize the source, including mass and spin, we see that neglecting terms $T_{\mu\nu}(q)\propto q^2$ in Eq.~\eqref{eq:GenericEMT} corresponds to describing the physics of the object in the limit in which 
\begin{equation}
    q\mathcal{R}_n\sim \frac{\mathcal{R}_n}{r}\ll 1\ ,
\end{equation}
\textit{i.e.} at distances much larger than the characteristic length scales of the source. Therefore, by definition, including local terms in the EMT will modify the characteristics of the source but not the induced gravitational multipoles, defining a class of equivalence between different EMTs that share the same long-range behavior. 

Up to now we have discussed only the case in which the EMT in momentum space is analytic in the transferred momentum, meaning that its real-space version is a localized matter-energy distribution, making it possible to reduce to vacuum Einstein equations very far from the source. However, within our space-momentum formalism, one can even consider non-localized EMTs, corresponding to sources captured by non-analytic contributions in the momentum-space version of the EMT. In the next section, we will discuss the subtlety of defining gravitational multipoles in non-vacuum scenarios within our formalism, comparing our conclusions with~\cite{Mayerson:2022ekj}.

\subsection{Multipoles in non-vacuum spacetimes}\label{sec:nonVacuumSolutions}

Let us now discuss what happens for non-vacuum spacetimes, namely when the EMT in momentum space has non-analytic contributions in the transferred momentum. In this case, even at infinity, the Einstein equations for the linear perturbations are
\begin{equation}
    \Box\Big(h_{\mu\nu}(x)-\frac{1}{2}\eta_{\mu\nu}h(x)\Big)=\frac{\kappa}{4}T_{\mu\nu}(x)\ .
\end{equation}
To give an idea of what are the subtleties in this scenario, within our formalism, consider the case in which 
\begin{equation}
    T_{\mu\nu}(q)=m \, u_\mu u_\nu \tilde{F}_{2, 1}\frac{\Lambda^2}{q^2}\Big(-q\cdot S\cdot S\cdot q\Big)^2\ ,
\end{equation}
with $\tilde{F}_{2, 1}$ some numerical free parameter and $\Lambda$ a mass-dimension coupling. From dimensional analysis we see that $T_{\mu\nu}(x)\sim r^{-d-2}$, hence it represents a well defined EMT in position space since its integral over $\mathbb{R}^d$ converges. However, the angular momentum induces an angular dependent piece in the metric  such as 
\begin{equation}
    h_{00}\sim G m \Lambda^2\frac{\rho}{r^2} \Big(n\cdot S\cdot S\cdot n\Big)^2\ ,
\end{equation}
where $n^{i}=x^i/r$.
From Eq.~\eqref{eq:MultipoleExpandedMetric} one can see that such metric is not in ACMC coordinates and, since $\tilde{F}_{2, 1}$ cannot be eliminated by a coordinate transformation, it means that for such source the harmonic gauge is not ACMC, and so gravitational multipoles cannot be extracted \`a la Thorne. Indeed, as shown in~\cite{Mayerson:2022ekj}, in non-vacuum environment the equivalence between harmonic and ACMC gauge, as well as the possibility to define gravitational multipoles, is not granted, and it depends on the fall-off of the non-localized EMT.

However, it is possible to define non-localized EMTs that give rise to spacetimes allowing for an ACMC reference frame.
Let us discuss an example considering as extra scale in the EMT an electric charge $Q$. Limiting ourselves to a static source, modulo local terms, we can write the most generic EMT as 
\begin{equation}\label{eq:ChargedEMT}
    T_{\mu\nu}(q)=m\ u_{\mu}u_\nu +Q^2\ q \ \Bigg(F^{(Q)}_{1, 1} \  u_{\mu}u_{\nu}+F^{(Q)}_{1, 2}\ \left(\eta_{\mu\nu}-\frac{q_\mu q_\nu}{q^2}\right)\Bigg)\ ,
\end{equation}
where $q^\mu T_{\mu\nu}(q)=0$. Computing the metric generated by the charged EMT in the harmonic gauge, we get
\begin{equation}\label{eq:ChargedMetric}
    \begin{aligned}
        h_{00}(x)&=\frac{2Gm}{r}+\Big(F^{(Q)}_{1, 1}+F^{(Q)}_{1, 2}\Big)\frac{4GQ^2}{\pi r^2}\ , \\
        h_{ij}(x)&=\frac{2Gm}{r}\delta_{ij}+\Big(4F^{(Q)}_{1, 2}x_ix_j\\
        &+(F^{(Q)}_{1, 1}-3F^{(Q)}_{1, 2})r^2\delta_{ij}\Big)\frac{4GQ^2}{\pi r^4}\ .
    \end{aligned}
\end{equation}
Since this is an ACMC coordinate system, we can read directly from Eq.~\eqref{eq:ChargedMetric} the gravitational multipoles induced by such source. Recalling Eq.~\eqref{eq:MultipoleExpandedMetric}, it is clear that the charged part of \eqref{eq:ChargedMetric} does not contribute to the gravitational multipoles, hence the odd-power in $q$ contributions in Eq.~\eqref{eq:ChargedEMT} are subleading with respect to terms considered in Eq.~\eqref{eq:ConservedEMT}. For the sake of completeness we notice that  for the choice of the coefficients
\begin{equation}
    F^{(Q)}_{1, 1}=-\frac{3}{64}\ , \qquad F^{(Q)}_{1, 2}=-\frac{1}{64}
\end{equation}
the charged EMT under consideration is the source of the Reissner-N\"ordstrom metric in the linearized approximation.

This example can indeed be generalized to all orders in $G$, since it is known that the Reissner-N\"ordstrom metric has all vanishing multipoles, meaning that every long-range contribution induced by the electric charge is sub-leading compared to the Schwarzschild metric. The same happens for Kerr-Newman metrics, which has the same gravitational multipole moments of the Kerr solution~\cite{Sotiriou:2004ud}. 

To summarize, non-localized EMTs can spoil the possibility of defining gravitational multipoles, and the treatment of non-vacuum spacetimes within this context can be subtle. However, if the EMT has a sufficiently rapid fall-off, it is still possible to define gravitational multipoles. In the latter scenario, special choices of non-localized EMT contributions do not modify the multipoles of the localized source, as in the Kerr-Newman case.

\section{Application to spinning BHs}\label{sec:ATSBHs}

BHs are among the most interesting solutions of Einstein equations. In addition to the presence of an event horizon, one of the most remarkable properties of BHs in four-dimensional GR is their uniqueness. It is indeed known that in $d=3$ the only rotating solution with a horizon is the Kerr metric, which happens to be also axisymmetric~\cite{Hawking:1973uf}. 
In higher dimensions the situation is much more complicated, and even if it has been proved that a stationary, non-extremal, asymptotically flat, rotating BH must admit at least one Killing vector that generates rotations~\cite{Hollands:2006rj}, there are no general uniqueness theorems for higher dimensional BHs in GR (see~\cite{Emparan:2008eg} for a review).

An explicit example of how such uniqueness is violated in higher dimensions occurs in $d=4$, for which in a specific parameter space region of mass and angular momentum there exist two different black solutions with different horizon topologies: Myers-Perry BHs with an $S^3$ horizon~\cite{Myers:1986un}, and black rings with a $S^2\times S^1$ horizon~\cite{Emparan:2001wn}. Even if it has been speculated that a uniqueness theorem can be restored in $d>3$ taking into account the stability of the solutions~\cite{Kol:2002dr}, there are no evidences up to date for such scenario.

However, some results can be obtained ensuring a sufficient amount of symmetries, and in $d=4$ if one assumes the existence of two rotational symmetries then it can be shown that the Myers-Perry solution is the unique stationary, non-extremal,
asymptotically flat, vacuum BH solution with spherical topology~\cite{Morisawa:2004tc}.
Despite the importance of such solutions, a comprehensive study of their multipolar structure is missing in the literature due to the difficulty to deal with the multipole definition in arbitrary dimensions. Our goal then is to employ the formalism developed so far in order to define the gravitational multipoles of the Myers-Perry solutions (which correspond to those of Kerr for $d=3$). In the following section we will firstly review the multipole structure of the Kerr metric using our formalism, and then we will discuss the multipole expansion of the Myers-Perry solution in $d=4$. Finally we will extend such construction to arbitrary dimension. 

\subsection{The Kerr case}

Let us consider the metric obtained by the EMT in Eq.~\eqref{eq:GenericEMT} for $d=3$. Since the $G_{n, m}$ factors do not contribute to multipoles, we can get rid of them without loss of generality and describe the long-range metric only by means of the form factors. Then, in order to fix the $F_{n, m}$ coefficients to recover the Kerr metric, we have to match Eq.~\eqref{eq:MetricFromEMT} with the linearized Kerr metric in harmonic coordinates expanded order by order in the spin (see Appendix \ref{App:KerrInHarm}). Moving to a reference frame in which the $z$-axis is oriented along the rotational symmetry, the spin tensor reduces to 
\begin{equation}
 S_{ij}=\begin{pNiceMatrix}[columns-width=auto]
        0 & a & 0 \\
        -a & 0 & 0 \\
        0 & 0 & 0
    \end{pNiceMatrix}\ ,
\end{equation}
and we managed to perform this matching up to the seventh order in the spin $a$. This requires 
\begin{equation}\label{eq:KerrFixing}
\begin{gathered}
    F_{0, 2}+F_{0, 1}=1\ ,\quad F_{2, 2}+F_{2, 1}=-\frac{1}{2}\ ,\quad F_{4, 2}+F_{4, 1}=\frac{1}{24}\ ,\\ F_{6, 2}+F_{6, 1}=-\frac{1}{720}\ ,\quad
    F_{1, 3}=1\ , \quad F_{3, 3}=-\frac{1}{6}\ ,\\ \quad F_{5, 3}=\frac{1}{120}\ ,\qquad F_{7, 3}=-\frac{1}{5040}\ .
\end{gathered}
\end{equation}
As we can see, while the current moments are uniquely fixed, there is a degeneracy between $F_{2\ell, 1}$ and $F_{2\ell, 2}$, due to the fact that we are in $d=3$, as already stressed in the previous section. 

Moreover, although we managed to perform this matching up to $O(a^7)$, it is possible to make an ansatz for the series at every order in spin as 
\begin{equation}\label{eq:d3Series}
    F_{2\ell, 2}+F_{2\ell, 1}=\frac{(-1)^\ell}{(2\ell)!}\ ,\qquad F_{2\ell+1, 3}=\frac{(-1)^\ell}{(2\ell+1)!}\ .
\end{equation}
Such series can be resummed introducing a dummy variable $\zeta$, and it leads to
\begin{equation}\label{eq:d3FormFactor}
\begin{aligned}
    F_{2}^{(d=3)}(\zeta)+F_{1}^{(d=3)}(\zeta)&=\sum_{\ell=0}^{+\infty}\Big(F_{2\ell, 2}+F_{2\ell, 1}\Big)\zeta^{2\ell}=\cos\zeta\ , \\
    F_{3}^{(d=3)}(\zeta)&=\sum_{\ell=0}^{+\infty}F_{2\ell+1, 3}\zeta^{2\ell}=\frac{\sin\zeta}{\zeta}\ .
\end{aligned}
\end{equation}
The form factor coefficients then can be extracted from the series expansion around $\zeta=0$ of such expressions, that generates only even power of the dummy variable. 
Moreover, for later purposes, it is important to notice that the resummed form factors can indeed be expressed in terms of spherical Bessel functions $j_n(\zeta)$ as 
\begin{equation}\label{eq:d3SphericalB}
\begin{gathered}
    F_{3}^{(d=3)}(\zeta)=j_0(\zeta)\ , \\ F_{2}^{(d=3)}(\zeta)+F_{1}^{(d=3)}(\zeta)=j_0(\zeta)-\zeta\, j_1(\zeta)\ .
\end{gathered}
\end{equation}
Interestingly the resummed form factors resemble the expressions of the Fourier Transform of the metric in the Kerr-Schild gauge that appear in the scattering amplitudes of~\cite{Bianchi:2023lrg}.

Finally, Eq.~\eqref{eq:d3Series} can be used in Eq.~\eqref{eq:d3Multipoles}, providing some closed form relations for the well-known infinite multipole towers of the Kerr metric as~\cite{Hansen:1974zz}
\begin{equation}\label{eq:KerrMultipoles}
    \begin{aligned}
        \mathbb{M}^{(2\ell)}_{A_{2\ell}}\Big|_{d=3}&=\frac{(4\ell-1)!!}{(2\ell)!}s_{a_1}\cdots s_{a_{2\ell}}\Big|_{TF}\ , \\         \mathbb{J}^{(2\ell+1)}_{i,A_{2\ell+1}}\Big|_{d=3}&=\frac{(4\ell)!!}{(2\ell+1)!}\ \epsilon_{ia_1k}s_{k}s_{a_2}\cdots s_{a_{2\ell}}\Big|_{TF}\ , \\
         \mathbb{G}^{(2\ell)}_{ij,A_{2\ell}}\Big|_{d=3}&=0\ .
    \end{aligned}
\end{equation}

\subsection{Myers-Perry in $d=4$}

We can repeat the argument of the previous subsection for the Myers-Perry solution in $d=4$. Following the procedure described in the Appendix~\ref{App:MPinHarm}, it is possible to express such linearized metric in harmonic gauge in order to match it with \eqref{eq:MetricFromEMT} by fixing the form factors. Moving to Myers-Perry coordinates $(y_1, x_1, y_2, x_2)$, in which the rotational axes are perpendicular to the planes $(y_1, x_1)$ and $(y_2, x_2)$, we can express the spin tensor as
\begin{equation}
 S_{ij}=\begin{pNiceMatrix}[columns-width=auto]
        0 & a_1 & 0 & 0 \\
        -a_1 & 0 & 0 & 0 \\
        0 & 0 & 0 & a_2 \\
        0 & 0 & -a_2 & 0
    \end{pNiceMatrix}\ ,
\end{equation}
from which we were able to perform such matching up to the seventh order in the spin. The form factors then read
\begin{equation}\label{eq:ExplicitMPFF}
    \begin{gathered}
      F_{0, 1}=1\ ,   \quad F_{2, 1}=-\frac{15}{32}\ ,\quad F_{4, 1}=\frac{63}{1024}\ ,\\
      F_{6, 1}=-\frac{243}{65536}\ ,\quad
      F_{0, 2}=0\ , \quad  F_{2, 2}=-\frac{3}{16}\ ,\\
      F_{4, 2}=\frac{9}{256}\ ,\quad F_{6, 2}=-\frac{81}{32768}\ ,\quad
        F_{1, 3}=1\ ,\\
        F_{3, 3}=-\frac{9}{32}\ ,\quad F_{5, 3}=\frac{27}{1024}\ ,\quad F_{7, 1}=-\frac{81}{65536}\ .
    \end{gathered}
\end{equation}

As we did in the previous subsection, we can make an ansatz for the series of form factors for every spin order. Indeed, one can verify that the above sequence can be reproduced by 
\begin{equation}\label{eq:d4Series}
\begin{gathered}
        F_{2\ell+2, 2}=-\frac{2}{3}\frac{(-1)^\ell}{(\ell)!(\ell+2)!}\left(\frac{3}{4}\right)^{2\ell+2}\ ,\\F_{2\ell+1, 3}=\frac{4}{3}\frac{(-1)^\ell}{(\ell)!(\ell+1)!}\left(\frac{3}{4}\right)^{2\ell+1}\ ,\\
        F_{2\ell,1}= F_{2\ell,2}+F_{2\ell+1,3}\ ,
\end{gathered}
\end{equation}
from which, using again a dummy variable $\zeta$, the form factors can be resummed as
\begin{equation}\label{eq:d4FormFactors}
\begin{aligned}
    F_{2}^{(d=4)}(\zeta)&=-\frac{2}{3}J_{2}\left(\frac{3}{2}\zeta\right)\ ,\\
    F_{3}^{(d=4)}(\zeta)&=\frac{4}{3\zeta}J_{1}\left(\frac{3}{2}\zeta\right)\ ,\\
    F_{1}^{(d=4)}(\zeta)&= F_{2}^{(d=4)}(\zeta)+ F_{3}^{(d=4)}(\zeta)\ ,
\end{aligned}
\end{equation}
where $J_\alpha(\zeta)$ is the Bessel function of the first kind. 
Once again these expressions resemble the ones that appear in the higher dimensional scattering amplitudes in the Kerr-Schild gauge~\cite{WorkInProgress_1}.

However, from \eqref{eq:d4FormFactors}, a recurrent structure with respect to the $d=3$ case can be noticed. Indeed, defining
\begin{equation}
    \mathcal{Z}_n^{(d)}(\zeta)=\Omega(d)\ \zeta^{-\frac{d-2}{2}}J_{n+\frac{d-2}{2}}\left(\frac{d-1}{2}\zeta\right)\ ,
\end{equation}
where 
\begin{equation}
    \Omega(d)=\frac{\Gamma(d/2)}{2^{2-d}(d-1)^{\frac{d-2}{2}}}
\end{equation}
is just a normalization factor chosen such that ${\mathcal{Z}_{n}^{(d)}(0)=1}$, it is possible to express the form factors as\footnote{Notice that the functions $\mathcal{Z}_n^{(d)}$ can be related to the spherical Bessel functions in arbitrary dimension.}
\begin{equation}\label{eq:d4SphericalB}
     F_{2}^{(d=4)}(\zeta)=-\frac{1}{2}\zeta\, \mathcal{Z}_1^{(d=4)}(\zeta)\ , \qquad
    F_{3}^{(d=4)}(\zeta)=\mathcal{Z}_0^{(d=4)}(\zeta)\ .
\end{equation}

Finally, using Eq.~\eqref{eq:d4Series} into \eqref{eq:GravitationalMultipoles}, we can give the complete series of the Myers-Perry gravitational multipoles in $d=4$ as
\begin{equation}
    \begin{aligned}
        \mathbb{M}^{(2\ell+2)}_{A_{2\ell+2}}\Big|_{d=4}&=\frac{(4+4\ell)!!}{(\ell+1)!^2}\left(\frac{3}{4}\right)^{2\ell+2}(-S\cdot S)_{A_{2\ell+2}}\Big|_{\rm STF}\ , \\
         \mathbb{J}^{(2\ell+1)}_{i,A_{2\ell+1}}\Big|_{d=4}&=\frac{1}{2}\frac{(2+4\ell)!!}{\ell!(\ell+1)!} \left(\frac{3}{4}\right)^{2\ell} \ S_{ia_1}(-S\cdot S)_{A_{2\ell}}|_{\rm ASTF}\ , \\
         \mathbb{G}^{(2\ell+2)}_{ij,A_{2\ell+2}}\Big|_{d=4}&=\frac{(4+4\ell)!!}{\ell!(\ell+2)!} \left(\frac{3}{4}\right)^{2\ell+2} \ S_{ia_1}S_{ja_2}(-S\cdot S)_{A_{2\ell}}|_{\rm RSTF}\ .
    \end{aligned}
\end{equation}
While mass and current multipoles were already found in~\cite{Heynen:2023sin}, we derived for the first time the complete tower of stress multipoles associated to the Myers-Perry solution in $d=4$.

\subsection{Myers-Perry in arbitrary dimensions}

From Eqs.~\eqref{eq:d3SphericalB} and \eqref{eq:d4SphericalB}, noticing the dependence on the spatial dimension in the resummed form factors, we conjecture that the form factors of the Myers-Perry solutions in arbitrary dimensions can be extracted as the coefficients of the series expansion of the following expressions
\begin{equation}\label{eq:dGenericSphericalB}
\begin{gathered}
     F_{2}^{(d)}(\zeta)=-\frac{1}{2}\zeta\, \mathcal{Z}_1^{(d)}(\zeta)\ , \qquad
    F_{3}^{(d)}(\zeta)=\mathcal{Z}_0^{(d)}(\zeta)\ ,\\
    F_{1}^{(d)}(\zeta)=F_{2}^{(d)}(\zeta)+F_{3}^{(d)}(\zeta)\ .
\end{gathered}
\end{equation}
As a sanity check, we can see that for $d=3$ the above definition reproduces Eq.~\eqref{eq:d3SphericalB}, 
\begin{equation}
\begin{aligned}
    &F_{1}^{(d=3)}(\zeta)+F_{2}^{(d=3)}(\zeta)=2F_{2}^{(d=3)}(\zeta)+F_{3}^{(d=3)}(\zeta)\\
    &=\mathcal{Z}_0^{(d=3)}(\zeta)-\zeta\, \mathcal{Z}_{1}^{(d=3)}(\zeta)\ .
    \end{aligned}
\end{equation}

Moreover, from \eqref{eq:dGenericSphericalB} we can extract the infinite series of Myers-Perry form factors in arbitrary dimensions as
\begin{equation}\label{eq:FFMParbitraryD}
\begin{aligned}
    F_{2\ell+2, 2}&=-\frac{1}{2}\Omega(d) \frac{(-1)^\ell}{\ell!\ \Gamma\left(\ell+2+\frac{d-2}{2}\right)}\left(\frac{d-1}{4}\right)^{2\ell+1+\frac{d-2}{2}}\ , \\
    F_{2\ell+1, 3}&=\Omega(d) \frac{(-1)^\ell}{\ell!\ \Gamma\left(\ell+1+\frac{d-2}{2}\right)}\left(\frac{d-1}{4}\right)^{2\ell+\frac{d-2}{2}}\ ,\\
    F_{2\ell+2, 1}&=F_{2\ell+2, 2}+F_{2\ell+1, 3}\ ,
\end{aligned}
\end{equation}
from which, replacing into Eq.\eqref{eq:GravitationalMultipoles}, we finally get the gravitational multipoles of Myers-Perry BHs in arbitrary dimensions
\begin{widetext}
\begin{equation}
    \begin{aligned}
        \mathbb{M}^{(2\ell+2)}_{A_{2\ell+2}}&=\frac{d-1}{4}\frac{(d+4\ell)!!\ (d+2\ell)}{(d-2)!!\ (\ell+1)!\ \Gamma\left(\ell+2+\frac{d-2}{2}\right)}\left(\frac{d-1}{4}\right)^{2\ell+1+\frac{d-2}{2}}(-S\cdot S)_{A_{2\ell+2}}\Big|_{\rm STF}\ , \\
         \mathbb{J}^{(2\ell+1)}_{i,A_{2\ell+1}}&=\frac{(d+4\ell-2)!!}{(d-2)!!\ \ell!\ \Gamma\left(\ell+1+\frac{d-2}{2}\right)}\left(\frac{d-1}{4}\right)^{2\ell+\frac{d-2}{2}} \ S_{ia_1}(-S\cdot S)_{A_{2\ell}}|_{\rm ASTF}\ , \\
         \mathbb{G}^{(2\ell+2)}_{ij,A_{2\ell+2}}&=\frac{d-1}{2}\frac{(d+4\ell)!!}{(d-2)!!\ \ell!\ \Gamma\left(\ell+2+\frac{d-2}{2}\right)}\left(\frac{d-1}{4}\right)^{2\ell+1+\frac{d-2}{2}} \ S_{ia_1}S_{ja_2}(-S\cdot S)_{A_{2\ell}}|_{\rm RSTF}\ .
    \end{aligned}
\end{equation}
\end{widetext}

Although the validity of Eq.~\eqref{eq:d3SphericalB} is conjectured, we managed to prove it also for $d=5$ up to $O(S^5)$. Indeed, by writing the Myers-Perry metric in ACMC coordinates (see Appendix \ref{App:MPinACMC}), generalizing the procedure outlined in~\cite{Heynen:2023sin}, we have verified that the above expressions hold in $d=5$.

\section{Matter source of Kerr and Myers-Perry BHs}\label{sec:MSKMP}

Once the explicit expression for the form factors that generates the Myers-Perry solution are found in Eq.~\eqref{eq:dGenericSphericalB}, an infinite number of EMTs can be defined such that they all share the same multipolar structure, differing between each other by local contributions. Among all the infinite choices of sources, in this section we want to compute the Fourier transform of the EMT in a specific configuration. Indeed, in analogy with the description of conserved charges in quantum field theory, form factors characterize the internal structure of the source, and considering them as constants coefficients corresponds to impose that the object we are describing is a point-like particle. 
In this spirit, in the following we show that the Israel EMT that sources the Kerr metric corresponds to a point-like distribution with all vanishing residual factors. By analogy we compute the EMT for the Myers-Perry solution in $d=4$ in the same setup. 

Moving to Myers-Perry coordinates in which the radial distance can be expressed in terms of Cartesian coordinates as
\begin{equation}
\begin{aligned}
r^2&=\sum_{k=1}^{\frac{d}{2}}(x_k^2+y_k^2)\quad \text{for} \quad d=\text{even}\ ,\\
r^2&=\sum_{k=1}^{\frac{d-1}{2}}(x_k^2+y_k^2)+z^2\quad \text{for} \quad d=\text{odd}\ ,
\end{aligned}
\end{equation}
each angular momenta is perpendicular to the plane $(y_k, x_k)$; then, we can set
\begin{equation}
    \zeta=\sqrt{-q\cdot S\cdot S\cdot q}=\sqrt{\sum_{k}q_{\perp, k}^{2}a_k^2}\ ,
\end{equation}
where the sum is performed over every angular momenta $a_k$ of the BH, and with ${q_{\perp, k}^2=q_{y_k}^2+q_{x_k}^2}$. We can now resum the EMT in Eq.~\eqref{eq:ConservedEMT}, from which we obtain\footnote{Notice that in the stress part of the EMT we have to consider an extra $\zeta^{-2}$ factor since the expansion starts at quadrupole order.}
\begin{equation}\label{eq:ConservedEMTresummed}
\begin{aligned}
    T^{\mu\nu}(q)&=m\ u^{\mu}u^{\nu}F_1^{(d)}(\zeta)+m \frac{{F}^{(d)}_{2}(\zeta)}{\zeta^2}(S\cdot q)^\mu (S\cdot q)^\nu\\
    &-\frac{i}{2}m\Big(u^\mu(S\cdot q)^\nu +u^\nu(S\cdot q)^\mu \Big){F}^{(d)}_{3}(\zeta)\ ,
\end{aligned}
\end{equation}
where the functions $F_n$ are defined in Eq.~\eqref{eq:dGenericSphericalB}. 
Then, considering the explicit form factor function that we found for the Myers-Perry solution, for $d>3$ one ends up with
\begin{equation}\label{eq:mpEMTresumDgeneric}
    \begin{aligned}
    &T^{00}(q)=m \left(\mathcal{Z}_{0}^{(d)}(\zeta)-\frac{1}{2}\zeta\ \mathcal{Z}_{1}^{(d)}(\zeta)\right)\ , \\
     &T^{0i}(q)=\frac{i}{2}m(S\cdot q)^i \mathcal{Z}_{0}^{(d)}(\zeta)\ ,\\
    &T^{ij}(q)=-\frac{m}{2}(S\cdot q)^i (S\cdot q)^j \frac{\mathcal{Z}_{1}^{(d)}(\zeta)}{\zeta}\ ,
    \end{aligned}
\end{equation}
while for $d=3$ we should consider the expression
\begin{equation}
    \begin{aligned}
    &T^{00}(q)\big|_{d=3}=m \left(\mathcal{Z}_{0}^{(d=3)}(\zeta)-F_{2}^{(d=3)}(\zeta)\right)\ , \\
    &T^{0i}(q)\big|_{d=3}=\frac{i}{2}m(S\cdot q)^i\mathcal{Z}_{0}^{(d=3)}(\zeta)\ ,\\
    &T^{ij}(q)\big|_{d=3}=m(S\cdot q)^i (S\cdot q)^j \frac{F_2^{(d=3)}(\zeta)}{\zeta^2}\ ,
    \end{aligned}
\end{equation}
where $F_2^{(d=3)}(\zeta)$ is just a gauge parameter that we can fix arbitrarily. 

However, both cases possess a cylindrical symmetry in momentum space, and we can set up the calculation of the EMT in a generic way. Indeed, from Eq.~\eqref{eq:MomentumEMT}, in the case of $d=\text{even}$ one gets
\begin{equation}\label{eq:EMTmsGenericD}
    \begin{aligned}
    T^{00}(x)&=m \prod_k\int_0^{+\infty}\frac{dq_{\perp, k}}{2\pi}q_{\perp, k}\\
    &\times J_0(q_{\perp, k}\rho_k)\Big(F_2^{(d)}(\zeta)+F_3^{(d)}(\zeta)\Big) \\
    T^{0i}(x)&=-\frac{1}{2}m(S\cdot \partial)^i\prod_k\int_0^{+\infty}\frac{dq_{\perp, k}}{2\pi}q_{\perp, k}J_0(q_{\perp, k}\rho_k)F_3^{(d)}(\zeta) \\
    T^{ij}(x)&=m(S\cdot \partial)^i (S\cdot \partial)^j \prod_k\int_0^{+\infty}\frac{dq_{\perp, k}}{2\pi}q_{\perp, k}\\
    &\times J_0(q_{\perp, k}\rho_k)\frac{F_2^{(d)}(\zeta)}{\zeta^2}\ ,
    \end{aligned}
\end{equation}
where $\rho_k^2=y_k^2+x_k^2$, while in the case of $d=\text{odd}$ we just need to add an extra $\delta(z)$ in front of each contribution.

\subsection{The Kerr case}

Let us now focus on the $d=3$ case. First of all, since there is only one angular momenta we have $\zeta=q_\perp a$, where we consider $a>0$ without loss of generality. Moreover, we can exploit the extra redundant gauge freedom\footnote{Notice that differently from the argument in Sec. \ref{sec:GaugeTransformation}, in this case $F_2^{(d=3)}(\zeta)$ is really a gauge parameter since does not give any local contributions.}, and move to a ``pressure-less gauge'' in which we set $F_2^{(d=3)}(\zeta)=0$. In this case, the EMT simply reads
\begin{equation}\label{eq:EMTkerrMP}
    \begin{aligned}
    &T^{00}(x)=m\  \delta(z)\int_0^{+\infty}\frac{dq_{\perp}}{2\pi}q_{\perp}J_0(q_{\perp}\rho)\cos(q_\perp a)\ , \\
    &T^{0i}(x)=-\frac{1}{2}m(S\cdot \partial)^i\delta(z)\int_0^{+\infty}\frac{dq_{\perp}}{2\pi}q_{\perp}J_0(q_{\perp}\rho)\frac{\sin(q_\perp a)}{q_\perp a}\ , \\
    &T^{ij}(x)=0\ .
    \end{aligned}
\end{equation}

Let us start with the computation of the mass density $T^{00}(x)$. Considering the following tabulated integral involving Bessel functions~\cite{Gradshteyn:1943cpj}
\begin{equation}
    \int_{0}^{+\infty}dz\ z \cos(c_1 z)J_0(c_2 z)=-\frac{c_1}{(c_1^2-c_2^2)^{3/2}}\Theta(c_1-c_2)\ ,
\end{equation}
where $\Theta(x)$ represents the Heaviside step function, we can compute the mass density energy as
\begin{equation}\label{eq:T00Kerr}
    T^{00}(x)=-\frac{m}{2\pi} \delta(z)\frac{a}{(a^2-\rho^2)^{3/2}}\Theta(a-\rho)\ .
\end{equation}
The energy density distribution turns out to be a thin equatorial disk with radius $a$. Moreover for $\rho=a$ there is a singularity that coincides with the curvature singularity of Kerr.
Indeed, Eq.~\eqref{eq:T00Kerr} is in perfect agreement with the original result of Israel~\cite{Israel:1970kp}  and with the derivation in~\cite{Balasin:1993kf}. However, while in the original work of Israel the shape of the energy distribution was assumed to be a disk of a finite radius, in~\cite{Balasin:1993kf} no assumption on the EMT was made, but still, in both derivations the full non-linear Kerr metric was employed. Remarkably, in the present derivation, instead, we did not make use of any assumptions on the EMT, except for requiring a vacuum solution of the Einstein equations and a source endowed with a mass and angular momentum.
Indeed, the only ingredient to derive Eq.~\eqref{eq:T00Kerr} is the particular multipolar structure of Kerr, which uniquely leads to its characteristic curvature (ring-)singularity.

It is important to notice that  the energy density is negative and hence violates the weak energy condition. We can indeed consider a static observer
with four-velocity ${U^\mu=(-1, 0, 0, 0)}$, from which 
\begin{equation}\label{eq:WEC}
    T_{\mu\nu}U^\mu U^\nu<0\ .
\end{equation}
This means that Eq.~\eqref{eq:T00Kerr} is not physical, even besides its singular nature. 

Moreover, due to the singularity at $\rho=a$, Eq.~\eqref{eq:T00Kerr} is defined up to a delta function. To get the total mass $m$ by integrating the mass density function we need to consider a regularized mass density function as
\begin{align}
    \tilde{T}^{00}(x)=&-\frac{m}{2\pi} \delta(z)\frac{a}{(a^2-\rho^2)^{3/2}}\Theta\Big(a(1-\epsilon)-\rho\Big)\nonumber\\
    &+\frac{m}{\sqrt{2\epsilon}}\delta(z)\frac{\delta(\rho-a)}{2\pi\rho}\ ,
\end{align}
such that one gets
\begin{equation}
   \lim_{\epsilon\rightarrow 0} \int d^3 x \tilde{T}^{00}(x)=m\ .
\end{equation}

Finally, in~\cite{Balasin:1993kf} it has been shown that \eqref{eq:T00Kerr} correctly reproduces the Schwarzschild limit for $a\rightarrow 0$. It is easy to see it in momentum space by looking at Eq.~\eqref{eq:EMTkerrMP}, while it is more subtle to see it in coordinate space due to the distributional nature of the expression. 

We can now move to the computation of the current part of the EMT. Since the $T^{0z}(x)$ component is vanishing, we need to compute only
\begin{equation}
\begin{aligned}
    T^{0x}(x)&=\frac{m}{4\pi\rho}y\, \delta(z)\int_0^{+\infty}dq_{\perp}q_{\perp}J_1(q_{\perp}\rho)\sin(q_\perp a)\ ,\\
    T^{0y}(x)&=-\frac{m}{4\pi\rho}x\, \delta(z)\int_0^{+\infty}dq_{\perp}q_{\perp}J_1(q_{\perp}\rho)\sin(q_\perp a)\ ,
\end{aligned}
\end{equation}
and using the known integral~\cite{Gradshteyn:1943cpj}
\begin{equation}
    \int_0^{+\infty}dz\ z \sin(c_1 z)J_1(c_2 z)=-\frac{c_2}{(c_1^2-c_2^2)^{3/2}} \Theta(c_1-c_2)\ ,
\end{equation}
one gets
\begin{equation}\label{eq:T0iKerr}
\begin{aligned}
    T^{0x}(x)&=-\frac{m}{4\pi} \delta(z)\frac{y}{(a^2-\rho^2)^{3/2}} \Theta(a-\rho)\ ,\\
    T^{0y}(x)&=+\frac{m}{4\pi} \delta(z)\frac{x}{(a^2-\rho^2)^{3/2}} \Theta(a-\rho)\ .
\end{aligned}
\end{equation}
\vspace{1cm}

Out of Eq.~\eqref{eq:T0iKerr} we can extract the angular momentum of the source by considering 
\begin{equation}
    \frac{d\vec{L}}{d\rho}=\int dz (2\pi \rho)\  \vec p\times \vec x\ ,
\end{equation}
where $p^i=T^{0i}(x)$ is the momentum density, leading to 
\begin{equation}\label{eq:AngMomKerr}
    dL_z=-\frac{m}{2}\delta(z)\frac{\rho^3}{(a^2-\rho^2)^{3/2}}d\rho \Theta(a-\rho)\ ,
\end{equation}
where $dL_x=dL_y=0$. This result matches exactly with what derived in~\cite{Israel:1970kp}.

Similar to the energy density term, even the current part of the EMT is non-physical, since for a stationary observer at infinity the angular velocity near the boundary of the disk can reach superluminal speed. Moreover, Eq.~\eqref{eq:AngMomKerr} needs to be regulated exactly as the energy density function. Indeed, by defining
\begin{equation}
\begin{aligned}
    d\tilde{L}_z&=-\frac{m}{2}\delta(z)\frac{\rho^3}{(a^2-\rho^2)^{3/2}}d\rho \Theta\Big(a(1-\epsilon)-\rho\Big)\\
    &+\frac{m a}{2\sqrt{2\epsilon}}\delta(z)\frac{\delta(\rho-a)}{2\pi \rho}\ ,
\end{aligned}
\end{equation}
one gets 
\begin{equation}
    \lim_{\epsilon\rightarrow 0}\int  d\tilde{L}_z= ma\ .
\end{equation}

\subsection{Myers-Perry in $d=4$}

Let us now focus on the EMT of Myers-Perry solutions in $d=4$. In this case there are two independent angular momenta\footnote{Notice that the normalization chosen for the angular momentum is different with respect to~\cite{Myers:1986un}. Indeed, we have normalized the spin densities at their ADM value, while in the original work of Myers and Perry they defined the spin parameters with a different normalization depending on the space-time dimension.} $a_1$ and $a_2$, and in the case in which $a_1\neq a_2$ the EMT reads
\begin{widetext}
\begin{equation}\label{eq:EMTmpd4}
    \begin{aligned}
    &T^{00}(x)=m \int_0^{+\infty}\frac{dq_{\perp, 1}}{2\pi}q_{\perp, 1}J_0(q_{\perp, 1}\rho_1)\int_0^{+\infty}\frac{dq_{\perp, 2}}{2\pi}q_{\perp, 2}J_0(q_{\perp, 2}\rho_2)\left(\frac{4}{3}\frac{J_1\left(\frac{3}{2}\zeta\right)}{\zeta}-\frac{2}{3}J_2\left(\frac{3}{2}\zeta\right)\right)\ , \\
    &T^{0i}(x)=-\frac{1}{2}m(S\cdot \partial)^i\int_0^{+\infty}\frac{dq_{\perp, 1}}{2\pi}q_{\perp, 1}J_0(q_{\perp, 1}\rho_1)\int_0^{+\infty}\frac{dq_{\perp, 2}}{2\pi}q_{\perp, 2}J_0(q_{\perp, 2}\rho_2)\left(\frac{4}{3}\frac{J_1\left(\frac{3}{2}\zeta\right)}{\zeta}\right)\ , \\
    &T^{ij}(x)=m(S\cdot \partial)^i (S\cdot \partial)^j \int_0^{+\infty}\frac{dq_{\perp, 1}}{2\pi}q_{\perp, 1}J_0(q_{\perp, 1}\rho_1)\int_0^{+\infty}\frac{dq_{\perp, 2}}{2\pi}q_{\perp, 2}J_0(q_{\perp, 2}\rho_2)\left(-\frac{2}{3}\frac{J_2\left(\frac{3}{2}\zeta\right)}{\zeta^2}\right)\ ,
    \end{aligned}
\end{equation}
\end{widetext}
where $\zeta=\sqrt{q_{\perp, 1}^2 a_1^2+q_{\perp, 2}^2 a_2^2}$.

Let us start the computation by considering the $T^{00}(x)$ component of the EMT. Using the Bessel identity 
\begin{equation}
    J_2(\tfrac{3}{2}\zeta)=\frac{4}{3}\frac{J_1(\tfrac{3}{2}\zeta)}{\zeta}-J_0(\tfrac{3}{2}\zeta)\ ,
\end{equation}
we can rewrite
\begin{equation}\label{eq:MPT00}
    T^{00}(x)=\frac{m}{(2\pi)^2}\Big(\frac{4}{9}A_1(x)+\frac{2}{3}A_0\Big)\ ,
\end{equation}
where 
\begin{widetext}
\begin{equation}\label{eq:Idefinition}
    \begin{aligned}
A_1&=\int_0^{+\infty}dq_{\perp, 1}q_{\perp, 1}J_0(q_{\perp, 1}\rho_1)\int_0^{+\infty}dq_{\perp, 2}q_{\perp, 2}J_0(q_{\perp, 2}\rho_2)\frac{J_1\left(\frac{3}{2}\zeta\right)}{\zeta}\ ,\\
A_0&=\int_0^{+\infty}dq_{\perp, 1}q_{\perp, 1}J_0(q_{\perp, 1}\rho_1)\int_0^{+\infty}dq_{\perp, 2}q_{\perp, 2}J_0(q_{\perp, 2}\rho_2)J_0\left(\frac{3}{2}\zeta\right)\ .
    \end{aligned}
\end{equation}

Considering then the Bessel integral~\cite{Gradshteyn:1943cpj}
\begin{equation}\label{eq:BesselSQRT}
    \int_0^{+\infty}dt\ J_{c_2}(\beta t)\frac{J_{c_1}\left(\alpha \sqrt{t^2+u^2}\right)}{\sqrt{(t^2+u^2)^{c_1}}}t^{c_2+1}=\frac{\beta^{c_2}}{\alpha^{c_1}}\Bigg(\frac{\sqrt{\alpha^2-\beta^2}}{u}\Bigg)^{c_1-c_2-1}J_{c_1-c_2-1}\left(u\sqrt{\alpha^2-\beta^2}\right)\Theta(\alpha-\beta)\ ,
\end{equation}
\end{widetext}
we get
\begin{equation}\label{Eq:I1}
    A_1=\frac{4}{3}\delta\Big(a_1^2\rho_2^2+a_2^2\rho_1^2-(\tfrac{3}{2}a_1a_2)^2\Big)\Theta(\tfrac{3}{2}a_1-\rho_1)\Theta(\tfrac{3}{2}a_2-\rho_2)\ ,
\end{equation}
where we used the relation 
\begin{equation}\label{eq:DeltaIdentity}
    \frac{\delta\left(\rho_2-\frac{a_2}{a_1}\sqrt{(\textstyle \frac{3}{2}a_1)^2-\rho_1^2}\right)}{2a_1^2\rho_2}=\delta\Big(a_1^2\rho_2^2+a_2^2\rho_1^2-(\textstyle \frac{3}{2}a_1a_2)^2\Big)\ ,
\end{equation}
and the Bessel ortogonality relation
from which 
\begin{equation}
    \int dx\ x\ J_p(c_1x)J_p(c_2x)=\frac{\delta(c_1-c_2)}{c_1}\ ,
\end{equation}
with $c_1, c_2, p\in \mathbb{Z}$.

Then, in order to compute $A_0$ and ensure ourselves to get a symmetric relation, we notice that Eq.~\eqref{eq:Idefinition} exhibits a bi-axial symmetry, and the result that one obtains integrating first in $q_{\perp, 1}$ (or $q_{\perp, 2}$) can be symmetrized in 
\begin{equation}\label{Eq:I0}
\begin{aligned}
    A_0&=\frac{1}{2}\Bigg(\frac{4}{3}\frac{\pi}{a_2}\delta(y_1)\delta(x_1)\delta(\tfrac{3}{2}a_2-\rho_2)\\
    &+\frac{4}{3}\frac{\pi}{a_1}\delta(y_2)\delta(x_2)\delta(\tfrac{3}{2}a_1-\rho_1)\Bigg)\ ,
\end{aligned}
\end{equation}
where we used some distribution property shown in Appendix \ref{app:Distributions}.
The resulting mass-energy distribution in Eq.~\eqref{eq:MPT00} is singular for $a_1^2\rho_2^2+a_2^2\rho_1^2=(\textstyle \frac{3}{2}a_1a_2)^2$, and vanishing everywhere else, describing a 3-ellipsoid embedded in $\mathbb{R}^4$ of semi-axis $\rho_1=\frac{3}{2}a_1$ and $\rho_2=\frac{3}{2}a_2$.
As a sanity check, it is easy to show that 
\begin{equation}\label{eq:MassNormalization}
    \int d\rho_1\, 2\pi\rho_1\int d\rho_2\, 2\pi\rho_2\ T^{00}=m\ .
\end{equation}

We can now compute the current part of the EMT in Eq.~\eqref{eq:EMTmpd4}. Since the integral is equal to the one already computed for $A_1$, we can already write
\begin{align}
    T^{0i}(x)=&-(S\cdot \partial)^i\frac{m}{(2\pi)^2}\frac{8}{9}\delta\Big(a_1^2\rho_2^2+a_2^2\rho_1^2-(\textstyle \frac{3}{2}a_1a_2)^2\Big)\nonumber\\
    &\times\Theta(\textstyle \frac{3}{2}a_1-\rho_1)\Theta(\textstyle \frac{3}{2}a_2-\rho_2)\ ,
\end{align}
and for specific components one gets
\begin{equation}\label{eq:EMTcurrentComponentsMPd4}
\begin{aligned}
    T^{0y_1}(x)&=\frac{x_1}{\rho_1}\frac{m}{(2\pi)^2}\frac{8}{9}\frac{\pi}{a_1}\delta(y_2)\delta(x_2)\delta(\textstyle \frac{3}{2}a_1-\rho_1)\ ,\\
     T^{0x_1}(x)&=-\frac{y_1}{\rho_1}\frac{m}{(2\pi)^2}\frac{8}{9}\frac{\pi}{a_1}\delta(y_2)\delta(x_2)\delta(\textstyle \frac{3}{2}a_1-\rho_1)\ ,
\end{aligned}
\end{equation}
and similar for $T^{0y_2}(x)$ and $T^{0x_2}(x)$. We can recover the angular momentum density tensor defined as
\begin{equation}
    l^{ij}=T^{0i}x^j-T^{0j}x^i\ ,
\end{equation}
where the angular momenta densities associated to the respective rotational planes can be identified as ${l_k=l^{y_kx_k}}$.
The explicit expressions then read 
\begin{equation}\label{eq:InfinitesimalAngMomSimplified}
\begin{aligned}
    l_1&=\frac{\rho_1}{a_1} \frac{m}{2\pi}\frac{4}{9}\delta(\textstyle\frac{3}{2}a_1-\rho_1)\delta(y_2)\delta(x_2)\ ,\\
    l_2&=\frac{\rho_2}{a_2} \frac{m}{2\pi}\frac{4}{9}\delta(\textstyle\frac{3}{2}a_2-\rho_2)\delta(y_1)\delta(x_1)\ .
\end{aligned}
\end{equation}
It is then easy to show that integrating such expressions one recovers the ADM angular momenta of the source
\begin{equation}
    L_k=\int d^4x\ l_k=a_k\, m\ .
\end{equation}

Finally we can focus on the stress part of the EMT. Limiting for simplicity to the study of
\begin{equation}
    P_k=T^{x_k x_k}+T^{y_k y_k}\ ,
\end{equation}
since the above expression only depends on the Laplacian, we can use the fact that the integrand does not depend on any angle, so that in 2 dimensions the radial part of the operator reads $\partial_{x_k}^2+\partial_{y_k}^2=\frac{1}{\rho_k}\partial_{\rho_k}\rho_k\partial_{\rho_k}$.
So then let us consider
\begin{equation}
\begin{aligned}
    P_1&=-a_1^2\frac{2}{3}\frac{m}{(2\pi)^2}\frac{1}{\rho_1}\partial_{\rho_1}\rho_1\partial_{\rho_1}\int_0^{+\infty}dq_2 \ q_2 J_0(\rho_2 q_2)\\
    &\times\int_0^{+\infty}dq_1 \ q_1 J_0(\rho_1 q_1)\frac{J_2(\zeta^2)}{\zeta^2}\ .
\end{aligned}
\end{equation}

Following the same steps as the previous calculation of the energy and current density, we can write
\begin{equation}
\begin{aligned}
    &P_1=\frac{m}{(2\pi)^2}\Bigg(\frac{1}{a_1}\frac{8}{9}\pi\delta(y_2)\delta(x_2)\delta(\tfrac{3}{2}a_1-\rho_1)\\
    &-\frac{32}{27}\delta\Big(a_1^2\rho_2^2+a_2^2\rho_1^2-(\tfrac{3}{2}a_1 a_2)^2\Big)\Theta(\tfrac{3}{2}a_1-\rho_1)\Theta(\tfrac{3}{2}a_2-\rho_2)\Bigg)\ ,
\end{aligned}
\end{equation}
and similar for $P_2$. We can also compute the integrated value, leading to 
\begin{equation}
    \int d^4x\ P_k=0\ .
\end{equation}
Moreover we can notice that $\delta_{ij}T^{ij}(x)=P_1(x)+P_2(x)$, from which considering Eq.~\eqref{eq:MassNormalization} we can make the following gauge invariant statement
\begin{equation}
    \int d^4x\, \eta_{\mu\nu}T^{\mu\nu}(x)=m\ .
\end{equation}

Finally, let us summarize some results. Eq.~\eqref{eq:MPT00} shows that the particle-like mass-energy density that sources a Myers-Perry BH in $d=4$ is exactly the curvature singularity of the full nonperturbative solution. Indeed, the 3-ellipsoid defined in Eq.~\eqref{eq:MPT00} is the hyper-surface in which the Myers-Perry solution is singular~\cite{Myers:1986un}. The same singularity structure is then present even in the current and stress part of the EMT. 

This is a quite interesting result since it shows that BH singularities are not a nonperturbative gravitational effect but they arise already at linearized level. Moreover they are intrinsically connected to the gravitational multipole expansion, establishing a surprising relationship between IR and UV BH physics. 
\\

\subsection{Singularity structure in arbitrary dimensions}

Looking at Eq.~\eqref{eq:EMTmsGenericD}, the generalization to an arbitrary number of dimensions should be straightforward. However, in the chosen setup of residual factors, higher-order Bessel functions make the possibility of deriving some analytical expression in terms of distribution less obvious. Restricting ourselves to the structure of the singularity though, we can still generalize it to an arbitrary number of dimensions and state that from Eq.~\eqref{Eq:I1}
\begin{widetext}
\begin{equation}\label{eq:SingularityD}
\begin{aligned}
T^{00}\Big|_{d=\text{even}}&\propto \frac{m}{(2\pi)^\frac{d}{2}}\frac{1}{\prod_k a_k^2}\delta\Big(\tfrac{\rho_k^2}{a_k^2}-(\tfrac{d-1}{2})^2\Big)\prod_k\Theta(\tfrac{d-1}{2}a_k-\rho_k)+\cdots\ ,\\
T^{00}\Big|_{d=\text{odd}}&\propto \frac{m}{(2\pi)^\frac{d-1}{2}}\frac{1}{\prod_k a_k^2}\delta(z)\delta\Big(\tfrac{\rho_k^2}{a_k^2}-(\tfrac{d-1}{2})^2\Big)\prod_k\Theta(\tfrac{d-1}{2}a_k-\rho_k)+\cdots\ ,
\end{aligned}
\end{equation}
\end{widetext}
where the ellipses stand for other contributions to the EMT, as for Eq.~\eqref{Eq:I0}.

We conclude by noting that Eq.~\eqref{eq:SingularityD} coincides exactly with the curvature singularity structure of Myers-Perry BHs in arbitrary spacetime dimensions~\cite{Myers:1986un}. This proves that the tight relationship between singularity and gravitational multipoles is a general feature independent on the number of dimensions.

\section{Conclusions}\label{sec:Conclusions}

Our study has provided new insights into the problem of identifying matter sources for spinning BHs within the framework of GR. By adopting a quantum field theory-inspired approach, we developed a momentum-space description of the EMT for a spinning point particle. This method allowed us to establish a direct mapping between the EMT and the infinite tower of gravitational multipole moments, simplifying the identification of the multipolar structure of BH spacetimes. Remarkably, this approach bypasses the intermediate step of computing the metric, offering a new and efficient way to characterize BH sources.

For the Kerr spacetime, we obtained a matter source corresponding to a thin, superluminally rotating disk originally derived by Israel. This agreement demonstrates the consistency of our framework with known results while shedding light on the intrinsic properties of BH singularities. Extending the analysis to higher-dimensional spacetimes, we derived the EMT for the Myers-Perry solution in five spacetime dimensions. The matter distribution in this case forms a three-dimensional ellipsoid, with additional stresses manifesting in higher-dimensional spaces.

A particularly striking result of this work is that the curvature singularity of BHs, traditionally considered a non-linear feature of GR, emerges already at the leading order in the gravitational coupling constant. This suggests that some key features of BH spacetimes, including their singularities, might have a simpler, linear origin when analyzed in momentum space. Additionally, our findings reveal a deep connection between the IR and UV regimes of BH physics, as the short-scale nature of the singularity is already captured by a resummation of the multipole structure at linear order in the coupling.

Overall, our momentum-space formalism provides a novel and versatile tool for studying the sources of spinning BHs and their generalizations. By establishing a clear link between the matter distribution and the BH multipole moments, our work lays the foundation for further exploration of other black objects in higher dimensions (such as black rings~\cite{Emparan:2001wn} and black saturns~\cite{Elvang:2007rd}), or horizonless compact objects. 
Future investigations could explore how this approach might extend to more realistic, non-pathological matter distributions, potentially resolving long-standing questions about the nature of BH interiors and the singularities they contain.
Restricting to four dimensions and to more phenomenological applications, our approach can be used to compute matter configurations that source the same multipolar structure as a Kerr BH but whose non-linear metric is regular and horizonless. Such solutions would be ideal BH mimickers~\cite{Cardoso:2019rvt}, since they share the exact multipoles of a Kerr BH and might be distinguished from the latter only at the horizon scale.

Furthermore, we focused on the case of EMT built out of the source mass and angular momentum, but our formalism can be generalized to account also for other \emph{intrinsic}, \textit{i.e.} not necessarily spin-induced, multipoles. The latter can arise for generically deformed compact objects, which might break the Kerr symmetries (e.g., current quadrupoles, mass and stress octupoles that break the equatorial symmetry, or generically moment tensors that break the axisymmetry)~\cite{Raposo:2018xkf,Raposo:2020yjy,Bena:2020see,Bianchi:2020bxa,Bena:2020uup,Bianchi:2020miz,Fransen:2022jtw}.

\begin{acknowledgments}
CG acknowledges the hospitality of the Universitat de Barcelona/ICCUB during the time this project was finalized.
This work is partially supported by Sapienza University of Rome (``Progetti per Avvio alla Ricerca - Tipo 1'',
protocol number AR1241906DC8FF32), the MUR PRIN Grant 2020KR4KN2 ``String Theory as a bridge between Gauge Theories and Quantum Gravity'', by the FARE programme (GW-NEXT, CUP:~B84I20000100001), and by the INFN TEONGRAV initiative.
\end{acknowledgments}

\appendix
\begin{widetext}
\section{Metric computation}\label{App:MetricComp}

Consider the linearized metric in Eq.~\eqref{eq:MetricFromEMT}. In order to compute explicitly the Fourier transform consider the master integral 
\begin{equation}
    \int \frac{d^dq}{(2\pi)^d} e^{-iq\cdot x}\frac{1}{q^2}=\frac{\rho}{4\pi}\ ,
\end{equation}
where $\rho$ is the harmonic function in higher dimension defined in Eq.~\eqref{Eq:RhoDefinition}. Then order by order in the spin expansion of the EMT, the relevant integral becomes
\begin{equation}
    \int \frac{d^dq}{(2\pi)^d} e^{-iq\cdot x}\frac{q_{i_1}\cdots q_{i_{\ell}}}{q^2}=(i)^{\ell}\partial_{i_1}\cdots\partial_{i_\ell}\frac{\rho}{4\pi}\ ,
\end{equation}
and replacing the above identity in Eq.~\eqref{eq:MetricFromEMT}, where the EMT is defined in \eqref{eq:GenericEMT}, for constant $F$ and $G$ factors one is able to compute the metric at arbitrary high order in the spin expansion. Organizing the spin expansion of the metric as 
\begin{equation}
h_{\mu\nu}=\sum_{\ell=0}^{+\infty}h_{\mu\nu}^{(\ell)}\ ,
\end{equation}
where $\ell$ labels the order of the spin $O(S^\ell)$ of each term,
for the sake of completeness we write in the following the first two non-trivial spin orders of the metric, namely the order $O(S^2)$
\begin{equation}\label{eq:S2Metric}
\begin{aligned}
    h_{00}^{(2)}(x)&=4\frac{d-2}{d-1}\frac{Gm\rho}{r^2}\Big((d-2) F_{21}+F_{22}\Big) \Big(d\, n\cdot S\cdot S\cdot n- S\cdot S\Big)\ ,\\
    h_{0i}^{(2)}(x)&=0\ ,\\
    h_{ij}^{(2)}(x)&=4\frac{d-2}{d-1}\frac{G m\rho}{r^2}\Bigg(-(d-1)\Big(d\, F_{22}(S\cdot n)_i(S\cdot n)_j+(F_{22}-2G_{21})(S\cdot S)_{ij}\\
    &-d\ G_{22}S\cdot S \ n_in_j+d\, G_{21}(n_i(S\cdot S\cdot n)_j+n_j(S\cdot S\cdot n)_i)\Big)\\
    &+\Big((F_{22}-F_{21}+G_{22}-d\, G_{22})S\cdot S+d(F_{21}-F_{22}) n\cdot S\cdot S\cdot n\Big)\delta_{ij}\Bigg)\ ,
\end{aligned}
\end{equation}
and the order $O(S^3)$
\begin{equation}\label{eq:S3Metric}
\begin{aligned}
    h_{00}^{(3)}(x)&=0\ ,\\
    h_{0i}^{(3)}(x)&=2(d-2)\frac{G m \rho}{r^3}dF_{33}\Big((d+2)(n\cdot S\cdot S\cdot n)(S\cdot n)_i-(S\cdot S)(S\cdot n)_i-2(S\cdot S\cdot S\cdot n)_i\Big)\ ,\\
    h_{ij}^{(3)}(x)&=0\ ,
\end{aligned}
\end{equation}
where $n^{i}=x^i/r$.

It is now straightforward to see how an infinitesimal gauge transformation acts on the metric. Indeed, considering the shift in Eq.~\eqref{eq:CoordTransf} and fixing the parameters as in \eqref{eq:GaugeFixingFFF}, one can sees that in Eqs.~\eqref{eq:S2Metric} and \eqref{eq:S3Metric} only the form factors are left.  Moreover, since the metric is expressed in harmonic gauge, in vacuum such reference frame belongs to the ACMC class~\cite{Mayerson:2022ekj}, and considering the normalization of the gravitational multipoles defined in Eq.~\eqref{eq:MultipoleExpandedMetric}, one can directly read the multipole moments out from the metric and recover \eqref{eq:GravitationalMultipoles}. 

Even though for simplicity we wrote here only the first two non-trivial spin orders of the metric, such calculation can be performed at arbitrary high order. We managed to compute the metric up to $O(S^7)$ in the spin expansion, from which we were able to infer the relation between form factors and gravitational multipoles at all orders as in Eq.~\eqref{eq:GravitationalMultipoles}.

\section{Kerr metric in harmonic gauge}\label{App:KerrInHarm}

Starting from the Kerr metric in Boyer-Lindquist coordinates
\begin{equation}
\begin{aligned}
    ds^2=&-\Big(1-\frac{2Gmr}{\Sigma}\Big)dt^2+\frac{\Sigma}{\Delta}dr^2+\Sigma d\theta^2+\Big(r^2+a^2+\frac{2Gmr}{\Sigma}a^2\sin^2\theta\Big)\sin^2\theta d\phi^2\\
    &-\frac{4Gmr}{\Sigma}a\sin^2\theta dtd\phi\ ,
\end{aligned}
\end{equation}
where 
\begin{equation}
    \Sigma=r^2+a^2\cos^2\theta\ ,\qquad\text{and}\qquad\Delta=r^2-2Gmr+a^2\ ,
\end{equation}
our goal here is to rewrite it in harmonic gauge.
Defining the spherical harmonic coordinates as $(T, R, \Theta, \Phi)$, and the associated Cartesian coordinates as
\begin{equation}
    \begin{cases}
    x=R\sin\Theta\cos\Phi\ ,\\
    y=R\sin\Theta\sin\Phi\ ,\\
    z=R\cos\Theta\ ,
    \end{cases}
\end{equation}
we can express the Kerr metric in harmonic coordinates by imposing 
\begin{equation}\label{eq:HarmonicBoxd3}
    g^{\mu\nu}D_\mu \partial_\nu (T, x, y, z)=0\ ,
\end{equation}
where each coordinate is treated as a scalar and $D_\mu$ is the covariant derivative. 
Since the Kerr metric is axisymmetric, we can define a coordinate transformation that does not involve the azimuthal coordinate as 
\begin{equation}\label{eq:HarmonicCoordTransfD4}
  T=t\ ,\qquad  R=r(R, \Theta)\ , \qquad \Theta=\theta(R, \Theta)\ ,\qquad \Phi=\phi \ . 
\end{equation}

The equation in \eqref{eq:HarmonicBoxd3} results in two independent partial differential equations and we can solve them by imposing an ansatz on the solutions expanding in the spin as
\begin{equation}
    r(R, \Theta)=R\sum_{i=0}^{n\text{PM}}\left(\frac{G m}{R}\right)^i\sum_{j=0}^{\lfloor \ell/2\rfloor}\left(\frac{a}{R}\right)^{2j}\sum_{k=0}^{j}\mathcal{C}_{i, 2j, k}^{(R)}P_{2k}(\cos\Theta)\ ,
\end{equation}
\begin{equation}\label{eq:ThetaD4HarmonicTransf}
    \cos\theta(R, \Theta)=\cos(\Theta)\sum_{i=0}^{n\text{PM}}\left(\frac{G m}{R}\right)^i\sum_{j=0}^{\lfloor \ell/2\rfloor}\left(\frac{a}{R}\right)^{2j}\sum_{k=0}^{j}\mathcal{C}_{i, 2j, k}^{(\Theta)}P_{2k}(\cos\Theta)\ ,
\end{equation}
where $P_n$ are the Legendre polynomials, $n$PM is the Post-Minkowksian order, $\ell$ represents the spin order and $\lfloor\cdot\rfloor$ stands for the integer part. The above ansatz is motivated to respect most of the symmetries of the Kerr metric. 

Moreover, after fixing the free coefficients to satisfy the harmonic gauge and the required symmetries, we are left with some gauge redundancies at every order in the spin expansion. Since at the end of the day we want to compare such metric with the one recovered from the generic EMT in Appendix \ref{App:MetricComp} in order to fix the form factors, we can get rid of those gauge redundancies that would be expressed in terms of the residual factors without any loss of generality. We managed to do such matching up to $O(S^7)$, and the resulting form factors are shown in Eq.~\eqref{eq:KerrFixing}.

\section{Myers-Perry metric in $d=4$ in harmonic gauge}\label{App:MPinHarm}

Let us consider the Myers-Perry solution in $d=4$
\begin{equation}\label{eq:MPmetric}
\begin{aligned}
ds^2=dt^2&-\frac{\mu}{\Sigma}\left(dt+\mathfrak{a}_1\, \sin^2\theta\, d\phi_1+\mathfrak{a}_2\, \cos^2\theta \, d\phi_2\right)^2-\frac{r^2\Sigma}{\Pi-\mu r^2}dr^2\\
&-\Sigma d\theta^2-(r^2+\mathfrak{a}_1^2)\sin^2\theta\, d\phi_1^2-(r^2+\mathfrak{a}_2^2)\cos^2\theta\, d\phi_2^2\ ,
\end{aligned}
\end{equation}
where 
\begin{equation}
\Sigma=r^2+\mathfrak{a}_1^2\cos^2\theta+\mathfrak{a}_2^2\sin^2\theta\ ,\qquad \Pi=(r^2+\mathfrak{a}_1^2)(r^2+\mathfrak{a}_2^2)\ , 
\end{equation}
and $\mathfrak{a}_1$ and $\mathfrak{a}_2$ are two independent spin parameters and 
\begin{equation}\label{eq:muParamMP}
    \mu=\frac{16 \pi G m}{(d-1)\Omega_{d-1}}
\end{equation}
with $\Omega_n$ the surface of a $n$-sphere.
Notice that the spin parameters in Eq.~\eqref{eq:MPmetric} are not the physical spin densities, and they are related to them through the following relation
\begin{equation}
    \mathfrak{a}_1=\frac{3}{2}a_1\qquad \text{and} \qquad \mathfrak{a}_2=\frac{3}{2}a_2\ .
\end{equation}

As we did in the $d=3$ case for the Kerr metric, in order to express the Myers-Perry metric in harmonic coordinates we can define a set of Cartesian harmonic coordinates as
\begin{equation}\label{eq:HarmCartesianD5}
    \begin{cases}
    x_1=R\sin\Theta\cos\Phi_1\ ,\\
    y_1=R\sin\Theta\sin\Phi_1\ ,\\
    x_2=R\cos\Theta\cos\Phi_1\ ,\\
    y_2=R\cos\Theta\sin\Phi_1\ ,
    \end{cases}
\end{equation}
related to the original coordinates through
\begin{equation}\label{eq:HarmonicCoordinatesD5}
  T=t\ ,\qquad  R=r(R, \Theta)\ , \qquad \Theta=\theta(R, \Theta)\ ,\qquad \Phi_1=\phi_1\ ,\qquad \Phi_2=\phi_2 \ , 
\end{equation}
and such that 
\begin{equation}\label{eq:BoxD5}
    g^{\mu\nu}D_\mu \partial_\nu  (T, x_1, y_1, x_2, y_2)=0\ .
\end{equation}
As it happens in the lower dimensional case, Eq.~\eqref{eq:BoxD5} results into two independent partial differential equations that we can solve by imposing an ansatz to the coordinate transformation perturbatively in the spins parameters. In this case the ansatz is much more involved and we can express it as
\begin{equation}
\begin{aligned}
    r(R, \Theta)&=R\sum_{i=0}^{n\text{PM}}(Gm\rho)^i \sum_{\sigma(p,q)}\mathcal{A}_{i}^{(p, q)}(\Theta)\ ,\\
    \cos \theta(R, \Theta)&=\cos(\Theta)\sum_{i=0}^{n\text{PM}}(Gm\rho)^i \sum_{\sigma(p,q)}\mathcal{B}_{i}^{(p, q)}(\Theta)\ ,
\end{aligned}
\end{equation}
where $\rho$ is defined in Eq.~\eqref{Eq:RhoDefinition} and
\begin{equation}
\begin{aligned}
    \mathcal{A}_{i}^{(p, q)}(\Theta)&=\sum_{k=0}^{n_k}\left(\frac{\mathfrak{a}_1^{p}\mathfrak{a}_2^{q}}{R^{p+q}}\right) \mathcal{C}_{i, p, q, 2k}^{(R)}P_{2k}\Big(f_{\sigma(p, q)}(\Theta)\Big)\ ,\\
     \mathcal{B}_{i}^{(p, q)}(\Theta)&=\sum_{k=0}^{n_k}\left(\frac{\mathfrak{a}_1^{p}\mathfrak{a}_2^{q}}{R^{p+q}}\right) \mathcal{C}_{i, p, q, 2k}^{(\Theta)}P_{2k}\Big(f_{\sigma(p, q)}(\Theta)\Big)\ .
\end{aligned}
\end{equation}

In the above coordinate transformation $p+q$ represents the order of the spin expansion, and up to it one has to consider every combination $\sigma(p, q)$ of the two spin parameters such that $p+q$ is always an even number and $f_{\sigma(p, q)}(\Theta)$ is defined as
\begin{equation}
    f_{\sigma(p, q)}\Theta=\begin{cases}
        \cos\Theta & p>q\ , \\
        \cos\Theta\sin\Theta & p=q\ ,\\
        \sin\Theta & p<q\ .
    \end{cases}
\end{equation}
To see explicitly how this works let us consider the expansion at second order in the spin $O(S^2)$. The terms we have to consider are $\mathfrak{a}_1^2, \mathfrak{a}_2^2$ and $\mathfrak{a}_1\mathfrak{a}_2$, which correspond respectively to $\sigma(p, q)=\{(2, 0), (0, 2), (1, 1)\}$. Next order in the spin expansion would be $O(S^4)$, which corresponds to $\sigma(p, q)=\{(4, 0), (3, 1),(2, 2), (1, 3), (0, 4)\}$, and so on. This ansatz is motivated to respect many of the symmetries of the Myers-Perry solution, even though is expressed in such a way that after the coordinate transformation some coefficients have to be fixed to respect all the properties of the solution. 

Then, as in the lower dimensional case, after the transformation the metric in harmonic coordinates will have some redundancies expressed in terms of free coefficients. Such parameters can be fixed in order to simplify the problem and do not spoil the match of the physical solution with the metric obtained from the EMT in \eqref{eq:GenericEMT} for $d=4$. Indeed, we were able to perform this procedure up to $O(S^7)$, leading to the unambiguous fixing of the form factors in Eq.~\eqref{eq:ExplicitMPFF}.

\section{Myers-Perry metric in $d=5$ in ACMC coordinates}\label{App:MPinACMC}

For an arbitrary number of even spacetime dimensions $D=d+1=2n+2$ with $n=1, 2, 3,...$, the Myers-Perry~\cite{Myers:1986un} metric reads
\begin{equation}\label{eq:MPODD}
    ds^2=-dt^2+\frac{\mu r}{\Pi F}\Big(dt+\sum_{k=1}^n \mathfrak{a}_k \mu_k^2 d\phi_k\Big)^2+\frac{\Pi F}{\Pi-\mu r}dr^2+\sum_{k=1}^n(r^2+\mathfrak{a}_k^2)(d\mu_k^2+\mu_k^2d\phi_k^2)+r^2d\alpha^2\ ,
\end{equation}
where 
\begin{equation}
    F=1-\sum_{k=1}^n\frac{\mathfrak{a}_k^2\mu_k^2}{r^2+\mathfrak{a}_k^2}\qquad \text{and} \qquad \Pi=\prod_{k=1}^n(r^2+\mathfrak{a}_k^2)\ ,
\end{equation}
with $\mathfrak{a}_k$ independent spin parameters and $\mu$ defined in Eq.~\eqref{eq:muParamMP}. Notice that the relation between the spin parameters and their ADM value is 
\begin{equation}
    \mathfrak{a}_k=\frac{d-1}{2}a_k\ .
\end{equation}
In the spirit of Sec.~\ref{Sec:GravMultFFO} we want to rewrite Eq.~\eqref{eq:MPODD} in generalized (higher dimensional) ACMC~\cite{Thorne:1980ru} coordinates and read the multipole moments for $d=5$. Starting from a set of spherical coordinates $(t, r, \theta, \psi, \phi_1, \phi_2)$
\begin{equation}
\begin{cases}
    \alpha=\cos\psi\ ,\\
    \mu_1=\sin\theta\sin\psi\ ,\\
    \mu_2=\cos\theta\sin\psi\ ,\\    
\end{cases}
\end{equation}
such that
\begin{equation}
\begin{cases}
    y_1=r\sin\theta\sin\psi\sin\phi_1\ ,\\
    x_1=r\sin\theta\sin\psi\cos\phi_1\ ,\\
    y_2=r\cos\theta\sin\psi\sin\phi_2\ ,\\
    x_2=r\cos\theta\sin\psi\cos\phi_2\ ,\\
    z=r\cos\psi\ ,
\end{cases}
\end{equation}
with $r^2=y_1^2+x_1^2+y_2^2+x_2^2+z^2$, $\phi_k$ polar angles and $\theta$ and $\psi$ azimuthal angles, expanding the metric in Eq.~\eqref{eq:MPODD} in powers of $G$ and angular momentum, we can see that is not in ACMC form, since for example
\begin{equation}
    g_{rr}=1-\frac{\Big(\mathfrak{a}_2^2\cos^2\theta+\mathfrak{a}_1^2\sin^2\theta\Big)\sin^2\psi}{r^2}+\cdots
\end{equation}
does not have the power dependence $\rho(r)/r^2$ to be a quadrupole term (see Eq.~\eqref{eq:MultipoleExpandedMetric}).

Our goal here is to find a coordinate transformation such that Eq.~\eqref{eq:MPODD} is expressed in ACMC coordinates, generalizing the procedure outlined in~\cite{Heynen:2023sin}. To this end we define the coordinates $(t, R, \Theta, \Psi, \phi_1, \phi_2)$ such that
\begin{equation}
\begin{aligned}
&r=R + \frac{1}{R} \left( -\frac{1}{4} \left( (\mathfrak{a}_1^2 + \mathfrak{a}_2^2) + (\mathfrak{a}_2^2 - \mathfrak{a}_1^2) \cos 2 \Theta \right) \sin^2\Psi \right)\\ 
&+ \frac{1}{R^3} \left( 
\frac{1}{4} \left( \mathfrak{a}_1^4 + \mathfrak{a}_2^4 + (\mathfrak{a}_2^4 - \mathfrak{a}_1^4) \cos2 \Theta \right) \sin^2\Psi 
- \frac{5}{32} \left( \mathfrak{a}_1^2 + \mathfrak{a}_2^2 + (\mathfrak{a}_2^2 - \mathfrak{a}_1^2) \cos 2 \Theta \right)^2 \sin^4\Psi 
\right)\ ,\\
&\cos\theta=\cos\Theta 
- \frac{(\mathfrak{a}_2^2 - \mathfrak{a}_1^2) \cos\Theta \sin^2\Theta}{2 R^2} \\
&+ \frac{(\mathfrak{a}_2^2 - \mathfrak{a}_1^2) \cos\Theta \sin^2\Theta}{16 R^4} 
\left( 
3 \mathfrak{a}_1^2 + \mathfrak{a}_2^2 
+ 5 (\mathfrak{a}_1^2 - \mathfrak{a}_2^2) \cos2\Theta
+ \left( 4 \mathfrak{a}_2^2 \cos^2\Theta + 4 \mathfrak{a}_1^2 \sin^2\Theta \right) \cos2\Psi 
\right)\ ,\\
&\cos\psi=\cos\Psi
+ \frac{1}{R^2} \left( 
\frac{1}{8} \left( \mathfrak{a}_1^2 + \mathfrak{a}_2^2 + (-\mathfrak{a}_1^2 + \mathfrak{a}_2^2) \cos2\Theta \right) 
\sin\Psi \sin2\Psi
\right) \\
&+ \frac{\sin\Psi}{R^4}
 \bigg( 
-\frac{1}{2} \left( 
\frac{1}{128} \left( 
11 \mathfrak{a}_1^4 - 14 \mathfrak{a}_1^2 \mathfrak{a}_2^2 + 11 \mathfrak{a}_2^4 
- 4 (\mathfrak{a}_1^4 - \mathfrak{a}_2^4) \cos2\Theta
- 7 (\mathfrak{a}_1^2 - \mathfrak{a}_2^2)^2 \cos4\Theta 
\right) \right) \sin2\Psi \\
&-\frac{1}{4} \left( 
\frac{7}{64} \left( \mathfrak{a}_1^2 + \mathfrak{a}_2^2 + (-\mathfrak{a}_1^2 + \mathfrak{a}_2^2) \cos2\Theta \right)^2 
\right) \sin4\Psi
\bigg) \ .
\end{aligned}
\end{equation}

With this coordinate transformation the Myers-Perry metric in $d=5$ is in ACMC-5, meaning that we can read multipoles moments up to $O(S^5)$. Then, extracting the gravitational multipoles in terms of form factors, one gets
\begin{equation}
    \begin{aligned}
      &F_{0, 1}=1\ ,   \qquad &F_{2, 1}=-\frac{3}{5}\ ,\qquad &F_{4, 1}=\frac{4}{35}\ ,\\
      &F_{0, 2}=0\ , \qquad  &F_{2, 2}=-\frac{1}{5}\ ,\qquad &F_{4, 2}=\frac{2}{35}\ ,\\
        &F_{1, 3}=1\ ,\qquad &F_{3, 3}=-\frac{2}{5}\ ,\qquad &F_{5, 3}=\frac{2}{35}\ .
    \end{aligned}
\end{equation}
Such form factors, extracted directly from the ACMC-5 version of the Myers-Perry metric in $d=5$, are in perfect agreement with Eq.~\eqref{eq:dGenericSphericalB} and in turn with \eqref{eq:FFMParbitraryD}. Notice that from the form factors one can easily obtain the gravitational multipoles using Eq.~\eqref{eq:GravitationalMultipoles}. Such procedure to obtain ACMC metrics can be generalized to higher dimensions, and more checks on our conjecture can be done. 

\section{Distributional evaluation of the EMT}\label{app:Distributions}

Let us compute the delta-function derivative that appears in the computation of the current and stress part of the EMT of the Myers-Perry solution in $d=4$. Considering the distribution function
\begin{equation}
    F(\rho_1, \rho_2)=\rho_1 \partial_{\rho_1}\Bigg\{\delta\Big( a_1^2\rho_2^2+a_2^2\rho_1^2-(\textstyle\frac{3}{2}a_1a_2)^2\Big)\Theta(\textstyle\frac{3}{2}a_1-\rho_1)\Theta(\textstyle\frac{3}{2}a_2-\rho_2)\Bigg\}\ ,
\end{equation}
in order to evaluate it we have to apply it to a test function $f(\rho_1, \rho_2)$
\begin{equation}
\begin{aligned}
    &\int d^4x F(\rho_1, \rho_2)f(\rho_1, \rho_2)\\
    &=\int d\rho_1 (2\pi\rho_1)\int d\rho_2 (2\pi\rho_2)\rho_1 \partial_{\rho_1}\Bigg\{\delta\Big( a_1^2\rho_2^2+a_2^2\rho_1^2-(\textstyle\frac{3}{2}a_1a_2)^2\Big)\Theta(\textstyle\frac{3}{2}a_1-\rho_1)\Theta(\textstyle\frac{3}{2}a_2-\rho_2)\Bigg\}f(\rho_1, \rho_2)\\
    &=\int d\rho_1 (2\pi\rho_1)\int d\rho_2 (2\pi\rho_2)\rho_1 \partial_{\rho_1}\Bigg\{\frac{\delta\Big(\rho_2-\textstyle\frac{a_2}{a_1}\sqrt{(\frac{3}{2}a_1)^2-\rho_1^2}\Big)}{2a_1^2\rho_2}\Theta(\textstyle\frac{3}{2}a_1-\rho_1)\Theta(\textstyle\frac{3}{2}a_2-\rho_2)\Bigg\}f(\rho_1, \rho_2)\ ,
\end{aligned}
\end{equation}
where we used Eq.\eqref{eq:DeltaIdentity}. Then we can integrate first over $\rho_2$ and then over $\rho_1$, leading to 
\begin{equation}\label{eq:DistributionOriginal}
    \begin{aligned}
        &2\pi \int d\rho_1 (2\pi\rho_1) \rho_1 \partial_{\rho_1}\Bigg\{\frac{1}{2a_1^2}\Theta(\textstyle\frac{3}{2}a_1-\rho_1)\Bigg\}f\Bigg(\rho_1, \textstyle\frac{a_2}{a_1}\sqrt{(\frac{3}{2}a_1)^2-\rho_1^2}\Bigg)\\
        &=-\frac{2\pi^2}{a_1^2} \int d\rho_1  \rho_1^2 \delta(\textstyle\frac{3}{2}a_1-\rho_1)f\Bigg(\rho_1, \textstyle\frac{a_2}{a_1}\sqrt{(\frac{3}{2}a_1)^2-\rho_1^2}\Bigg)\\
        &=-\frac{9\pi^2}{2}f(\textstyle\frac{3}{2}a_1, 0)\ .
    \end{aligned}
\end{equation}

We can show that the exact same result can be obtained by considering the distribution in Eq.~\eqref{eq:InfinitesimalAngMomSimplified}. Indeed considering the distribution
\begin{equation}
    \tilde{F}(\rho_1, \rho_2)=-\pi\frac{\rho_1}{a_1^2}\delta(\textstyle{\frac{3}{2}}a_1-\rho_1)\delta(y_2)\delta(x_2)\ ,
\end{equation}
one gets
\begin{equation}
\begin{aligned}
    &\int d^4x \tilde{F}(\rho_1, \rho_2)f(\rho_1, \rho_2)=-\pi\int d\rho_1(2\pi \rho_1)\frac{\rho_1}{a_1^2}\delta(\textstyle{\frac{3}{2}}a_1-\rho_1)f(\rho_1, 0)\\
    &=-2\pi^2\int d\rho_1\frac{\rho_1^2}{a_1^2} \delta({\tfrac{3}{2}}a_1-\rho_1)f(\rho_1, 0)=-\frac{9\pi^2}{2}f(\tfrac{3}{2}a_1, 0)\ ,
\end{aligned}
\end{equation}
which is exactly equivalent to Eq.~\eqref{eq:DistributionOriginal}. So then we can state the following relation in a distributional sense
\begin{equation}
    \rho_1 \partial_{\rho_1}\Bigg\{\delta\Big( a_1^2\rho_2^2+a_2^2\rho_1^2-(\tfrac{3}{2}a_1a_2)^2\Big)\Theta(\tfrac{3}{2}a_1-\rho_1)\Theta(\tfrac{3}{2}a_2-\rho_2)\Bigg\}=-\pi\frac{\rho_1}{a_1^2}\delta({\tfrac{3}{2}}a_1-\rho_1)\delta(y_2)\delta(x_2)\ .
\end{equation}

Operatively speaking, since the distribution is meant to be applied to a two variable test function, whenever it appears a derivative of the delta function together with two step functions, we can just commute the derivative with the delta-function and apply it to the step functions. This applies also in the case in which the derivative is performed with respect to the spins. 
\clearpage
\end{widetext}

\bibliography{biblio}

\end{document}